\documentclass[twocolumn]{aastex63}
\usepackage{graphicx}
\graphicspath{{figure/}}
\usepackage{hyperref}
\usepackage{apjfonts}
\usepackage{appendix}

\newcommand{\lum}{erg~s\ensuremath{^{-1}}}

\newcommand{\lbol}{\ensuremath{L\mathrm{_{bol}}}}

\newcommand{\eddratio}{\ensuremath{L\mathrm{_{bol}}/L\mathrm{_{edd}}}}
\newcommand{\lamedd}{\ensuremath{\lambda\mathrm{_{edd}}}}

\newcommand{\fc}{\ensuremath{L\mathrm{_{W1}}/L\mathrm{_{bol}}}}

\newcommand{\msun}{\ensuremath{M_{\odot}}}

\newcommand{\mbh}{\ensuremath{M_\mathrm{BH}}}
\newcommand{\wise}{\emph{WISE}}
\newcommand{\neowise}{\emph{NEOWISE}}

\shorttitle{MIR Periodic AGNs}
\shortauthors{Luo, Jiang \& Liu}

\submitjournal{\apj}
\received{2024 May 24}
\revised{2024 November 9}
\accepted{2024 November 11}

\begin{document}

\title{A Systematic Search for Candidate Supermassive Black Hole Binaries Using Periodic Mid-Infrared Light Curves of Active Galactic Nuclei}

\author[0009-0007-1153-8112]{Di Luo}
\affiliation{CAS Key Laboratory for Research in Galaxies and Cosmology, Department of Astronomy, University of Science and Technology of China, Hefei, 230026, China; jnac@ustc.edu.cn}
\affiliation{School of Physical Sciences, University of Science and Technology of China, Hefei, 230026, China}

\author[0000-0002-7152-3621]{Ning Jiang}
\affiliation{CAS Key Laboratory for Research in Galaxies and Cosmology, Department of Astronomy, University of Science and Technology of China, Hefei, 230026, China; jnac@ustc.edu.cn}
\affiliation{School of Astronomy and Space Sciences,
University of Science and Technology of China, Hefei, 230026, China}

\author[0000-0003-0049-5210]{Xin Liu}
\affiliation{Department of Astronomy, University of Illinois at Urbana-Champaign, Urbana, IL 61801, USA}
\affiliation{National Center for Supercomputing Applications, University of Illinois at Urbana-Champaign, Urbana, IL 61801, USA}
\affiliation{Center for Artificial Intelligence Innovation, University of Illinois at Urbana-Champaign, 1205 West Clark Street, Urbana, IL 61801, USA}

\begin{abstract}

Periodic variability in active galactic nuclei (AGNs) is a promising method for studying sub-parsec supermassive black hole binaries (SMBHBs), which are a challenging detection target. While extensive searches have been made in the optical, X-ray and gamma-ray bands, systematic infrared (IR) studies remain limited. Using data from the Wide-field Infrared Survey Explorer (WISE), which provides unique decade-long mid-IR light curves with a six-month cadence, we have conducted the first systematic search for SMBHB candidates based on IR periodicity. Analyzing a parent sample of 48,932 objects selected from about half a million AGNs, we have identified 28 candidate periodic AGNs with periods ranging from 1,268 to 2,437 days (in the observer frame) by fitting their WISE light curves with sinusoidal functions. However, our mock simulation of the parent sample indicates that stochastic variability can actually produce a similar number of periodic sources, underscoring the difficulty in robustly identifying real periodic signals with WISE light curves, given their current sampling. Notably, we found no overlap between our sample and optical periodic sources, which can be explained by a distinct preference for certain periods due to selection bias.  By combining archived data from different surveys, we have identified SDSS J140336.43+174136.1 as a candidate exhibiting periodic behavior in both optical and IR bands, a phenomenon that warrants further validation through observational tests. Our results highlight the potential of IR time-domain surveys, including future missions such as the Nancy Grace-Roman Space Telescope, for identifying periodic AGNs, but complementary tests are still needed to determine their physical origins such as SMBHBs.

\end{abstract}

\keywords{Active galactic nuclei (16); Infrared astronomy(786); Quasars (1319); Supermassive black holes (1663); Time domain astronomy (2109)}

\section{introduction}

Supermassive black holes (SMBHs) are intriguing celestial objects that reside at the centers of galaxies~\citep{KH2013}. When galaxies merge, it is anticipated that the central black holes will form binary systems known as SMBH binaries (SMBHBs,~\citealt{Begelman1980}). These binary systems offer valuable insights into various astrophysical phenomena, including galaxy formation~\citep{Colpi2011}, gravitational wave (GW) emission~\citep{Hughes2009}, and the evolution of SMBHs~\citep{Merritt2013}. More massive binaries are pulsar-timing array (PTA) sources (e.g.,~\citealt{Arzoumanian2018}), while less massive binaries are targeted by space-based experiments such as LISA~\citep{Klein2016}. They provide a laboratory to directly test strong-field general relativity~\citep{Hughes2009, Centrella2010}.

However, despite their theoretical significance, the direct electromagnetic detection of close SMBHBs below sub-parsec separation has proven to be a challenging endeavor (see recent reviews such as~\citealt{DeRosa2019, DOrazio2023}) while numerous SMBH pairs at larger scales have been identified through the presence of dual active galactic nuclei (AGNs)~(e.g., \citealt{Komossa2003, Zhou2004, Liu2010, Liu2011, Liu2013, Koss2011, Liu2018, Chen2022}). When a binary has exhausted its interactions with stars but has not approached close enough to emit significant gravitational radiation, its orbit does not have an obvious mechanism for decay. This long-standing challenge presents a major uncertainty in estimating the abundance of SMBHB mergers as GW sources. In theory, the bottleneck may be overcome in gaseous environments (e.g., ~\citealt{Gould2000, Cuadra2009, Chapon2013, delValle2015}), in triaxial or axisymmetric galaxies (e.g.,~\citealt{Khan2016, Kelley2017}), and/or by interacting with a third BH in hierarchical mergers (e.g.,~\citealt{Blaes2002, Kulkarni2012, Bonetti2018}).

A recent and promising approach for identifying potential close SMBHB candidates has emerged in the thriving field of time-domain astronomy, which focuses on analyzing the periodic light curves of AGNs. The periodicity observed in these light curves can arise from changes in the accretion rate onto the black holes~\citep{Graham2015b} or from the relativistic Doppler boost resulting from the highly relativistic motion of gas in the mini accretion disk around the smaller black hole in the binary system~\citep{DOrazio2015}. Extensive research has been conducted using optical light curves to search for SMBHBs, and the methods employed in these studies have become well-established~\citep{Graham2015a, Charisi2016, Zheng2016, Liu2019, Chen2020, Liao2021, Chen2024}. However, many of the known candidates have been shown to be subject to false positives due to stochastic quasar variability (e.g.,~\citealt{Vaughan2016}). Additionally, recent findings suggest that some AGNs may display periodic variations only at some specific epochs (e.g.,~\citealt{Jiang2022, ONeill2022}) further enhance the difficulties of periodic searches. Furthermore, previous surveys were only sensitive to the most massive quasars at high redshift ($z\gtrsim 2$) which should have already gone through their major merger process (e.g.,~\citealt{Shen2009}). Regardless of the process that imprints periodicity, it is generally accepted that the timescale is primarily determined by the binary orbital period. SMBHBs with masses between $10^6-10^{10}$~\msun\ and separations of 0.01 pc possess orbital periods that span from years to several decades, making them possibly detectable by current time-domain surveys. In addition to the optical band, the periodic searches have also been conducted in other wavelength regimes, such as in X-ray~\citep{Serafinelli2020, Liu2020} and gamma-ray bands~\citep{Sandrinelli2018, Holgado2018}.

Periodic infrared (IR) light curves are also expected as a result of reverberation mapping of the circumbinary dusty torus in periodic AGNs. However, the search for SMBHBs using IR data has not been extensively explored, despite the detection of an IR time lag in an individual SMBHB candidate~\citep{Jun2015}, which highlights the potential of utilizing periodic IR light curves for SMBHB identification~\citep{DOrazio2017}. The IR band not only provides an alternative approach to searching for SMBHBs but also offers distinct advantages. Firstly, IR photons are less susceptible to dust extinction, allowing them to penetrate through obscuring material, which is often abundant in galactic nuclear regions and galaxy merger systems. By extending the search to the IR wavelength range, we can uncover a population of SMBHB candidates that may have been previously overlooked, thereby expanding our knowledge of the prevalence and characteristics of these elusive systems. Even for unobscured AGNs, the variability amplitudes in the mid-IR band are typically larger compared to the optical band due to lower background contamination. Additionally, as discussed in~\citet{DOrazio2017}, the dust-echo model can assist in determining the origin of central periodicity and place constraints on the physical characteristics of SMBHBs and their dust torus. To address this, we perform the first systematic search for SMBHB candidates via periodic IR light curves. The data we use come from the Wide-field Infrared Survey Explorer (WISE, \citealt{Wright2010}) and its successor NEOWISE~\citep{Mainzer2014}, which have provided us with a public dataset from February 2010 to December 2023 with a half-year cadence except for a 2.5-year gap for each object. In this study, we primarily focus on sinusoidal periodic signals, which have been commonly adopted in modeling the optical light curves of SMBHB candidates (e.g.,~\citealt{DOrazio2015}), and on modeling the IR light curves using the dust-echo model~\citep{DOrazio2017}.

The paper is structured as follows. Section~\ref{data} introduces the data and methods employed to search for IR periodic AGNs. Section~\ref{results} presents the search and corresponding analysis results, where we show compelling evidence demonstrating the highly improbable nature of these periodic light curves being generated by random processes. Section~\ref{discuss} compares selected candidates with normal AGNs and with previous studies that utilized optical light curves. Additionally, we discuss SDSS J140336.43+174136.1 as a candidate exhibiting periodic behavior in both optical and IR bands. Finally, we summarize the main results and suggest directions for future work in Section~\ref{conclusion}. We assume a cosmology with $H_{0} =66.88$ km~s$^{-1}$~Mpc$^{-1}$, $\Omega_{m} = 0.32$, and $\Omega_{\Lambda} = 0.68$, adopted from~\citet{Planck2020}.

\section{Data and Methods}
\label{data}

\subsection{IR data}

The WISE was originally launched to conduct a full-sky survey in four MIR bands centered at 3.4, 4.6, 12, and 22~$\mu$m (labeled W1-W4) from February to August 2010~\citep{Wright2010}. The solid hydrogen cryogen used to cool the W3 and W4 instrumentation was depleted later and the spacecraft was placed in hibernation in 2011 February. However, it was reactivated in 2013 October under the new name \neowise-R, with only W1 and W2 operational, specifically to search for asteroids that could potentially pose a threat of impact on Earth~\citep{Mainzer2014}. The \wise\ and its new mission \neowise\ operate in the same manner, that is, scanning a specific sky area every six months, with an average of 12 or more individual exposures taken within each epoch, typically spanning a single day. Consequently, each target in the sky has been observed 22--23 times separated by a six-month interval up to December 2023.

The single-exposure photometry of WISE and NEOWISE are archived in the AllWISE Multiepoch Photometry Table and NEOWISE-R Single Exposure (L1b) Source Table~\footnote{https://irsa.ipac.caltech.edu/cgi-bin/Gator/nph-scan?mission=irsa\&submit=Select\&projshort=WISE}, respectively. The photometry we adopted is measured by point-spread function (PSF) profile fitting, which is suitable for quasars. We first binned the data within each epoch since the intraday IR variability is negligible except for very radio-loud AGNs~\citep{Jiang2012} and the anticipated periods we aim to detect are on much longer timescales. The variance of the binned data was recalculated as shown in Equation~\ref{er}:

\begin{equation}
[\delta mag(t_i)]^2 = \frac{1}{n_i -1} \sum_{j=1}^{n_i} [mag_j -mag(t_i)]^2 + \frac{1}{n_i} \sigma_{s.s.}^2
\label{er}
\end{equation}

where $n_i$ denotes the number of data points within the $i$th epoch, $mag(t_i)$ denotes the mean magnitude within the $i$th epoch, and $\sigma_j$ denotes the photometric uncertainty of the $j$th data point. The first part represents the contribution from photometric uncertainty and short-term variability, while the second part represents the contribution from the system stability of WISE. We adopt $\sigma_{s.s.} =0.029$ mag for \wise\ and $\sigma_{s.s.} =0.016$ mag for \neowise\ from~\citet{Lyu2019}.

Prior to binning, the data were filtered based on the quality flags in the catalog following~\citet{Jiang2021}. Moreover, outliers that fell outside the $3\sigma$ range of the single data point at each epoch were eliminated, using the mean magnitude and mean photometric uncertainty at that epoch as references. The mean and variance of the single data points were then re-computed, resulting in the binned data used for our analysis. This step was crucial to ensure the accuracy and reliability of the data used in our study.

\subsection{AGN Sample}
\label{sample}

The AGN sample we chose is the widely-used Million Quasar Catalog (v8, 2 August 2023;~\citealt{Flesch2023}). This is a compendium of 907,144 type-I QSOs and AGNs, largely complete from the literature up to 30 June 2023. 66,026 QSO candidates are also included, calculated via radio/X-ray association (including double radio lobes) as being 99\% likely to be quasars. Blazars and type-II objects are also included, bringing the total count to 1,021,800. To ensure the purity of our candidates, we only selected objects labeled as types 'Q', 'A', 'B', 'K', 'N' in the catalog (see what the legends of these types refer to in the note of Table~\ref{cat}), as these are confirmed AGNs. As a result, the number of objects amounts to 955,744. To enable a reliable long-term variability analysis, we apply a criterion that requires a minimum of 15 detected epochs (about 2/3 of all detected epochs) for each source. This left us with a subset of 576,260 AGNs.

In order to rationalize the computational effort in further analysis, we choose to analyze only AGNs with obvious variability. This selection strategy is crucial to ensure that candidates meet the criteria outlined in selection criterion 5 (see Section~\ref{sc}). Furthermore, light curves displaying subtle variability tend to pose a challenge in terms of model constraints within the DRW framework (see Section~\ref{stochastic}). To address this issue, we introduce a requirement that, for a given light curve, at least one data point must satisfy $mag(t_i)-2\delta mag(t_i)> mag_{mean}$ and at least one data point should satisfy $mag(t_i)+2\delta mag(t_i)< mag_{mean}$. This allows us to narrow down our sample to 53,496 AGNs, ensuring that the selected candidates have the necessary characteristics for further analysis. We also require that the light curves be well constrained by the DRW model (see section~\ref{stochastic}), resulting in a parent sample of 48,932 AGNs.

\subsection{Finding Periodicity}
\label{method}

\begin{figure*}
\centering
\begin{minipage}{1.0\textwidth}
\centering{\includegraphics[width=0.49\textwidth]{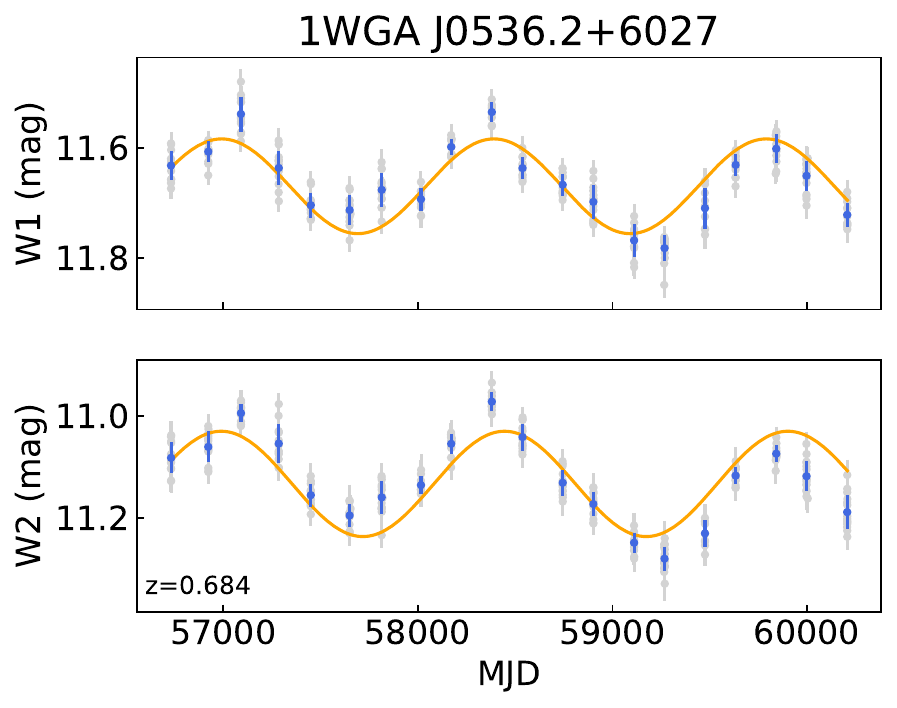}}
\label{lc1}
\centering{\includegraphics[width=0.49\textwidth]{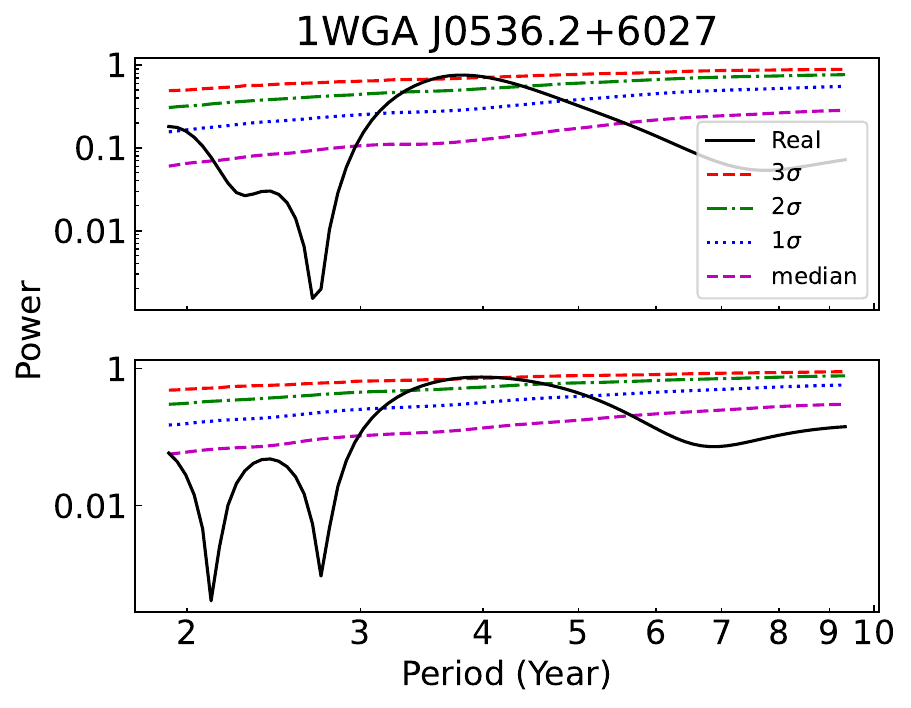}}
\label{pg1}
\end{minipage}
\begin{minipage}{1.0\textwidth}
\centering{\includegraphics[width=0.49\textwidth]{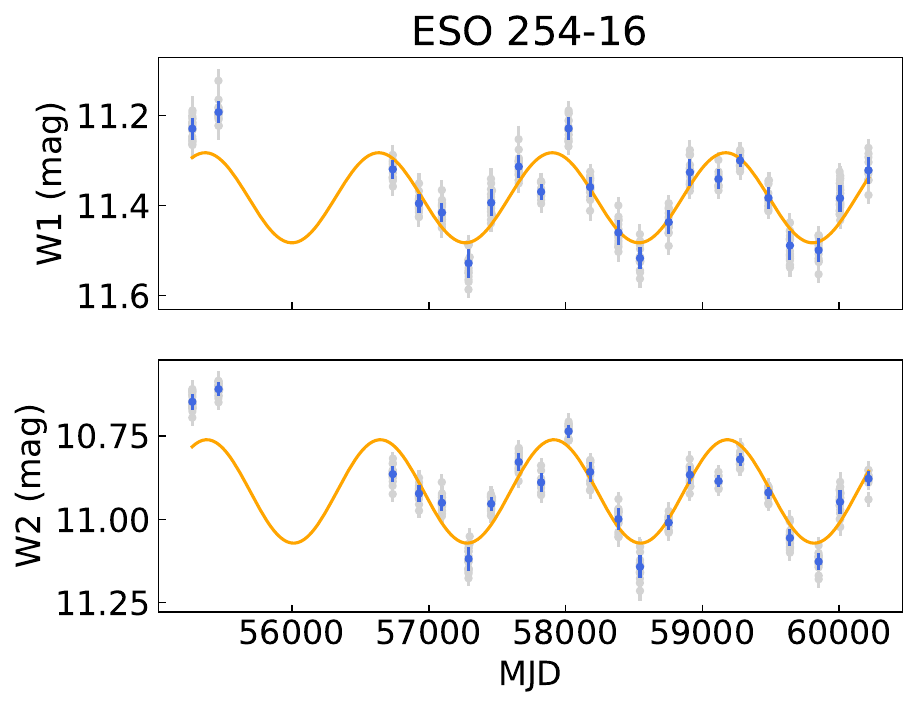}}
\label{lc2}
\centering{\includegraphics[width=0.49\textwidth]{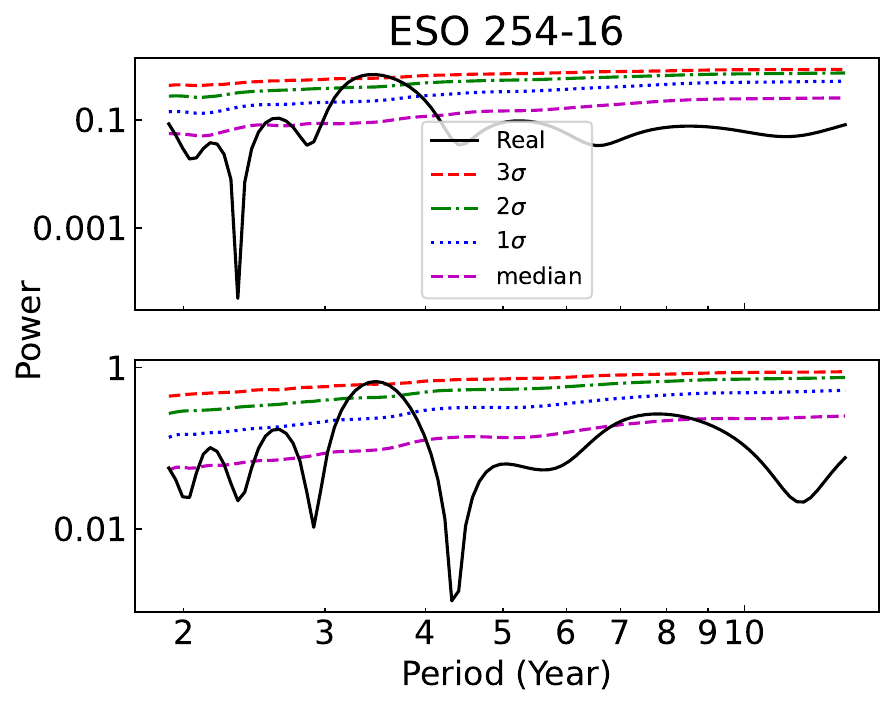}}
\label{pg2}
\end{minipage}
\caption{Two special examples of candidates. The top panel specifically highlights the highest redshift source at $z=0.684$ and the bottom panel highlights the candidate showing the most periods. Left: the grey error bars represent the original data (with outlier data points removed). The blue error bars represent the binned data. The orange sinusoids represent the best-fitting sine curves. Right: the black curves show the periodograms of candidates, while other curves show different significance level calculated from 100,000 simulated light curves.}
\label{lc}
\end{figure*}

To identify sources exhibiting periodic variability, we employed a sinusoidal function to fit the light curves of both W1 and W2 bands. The fitting process involved minimizing the following equation using the $\chi^2$ statistics as Equation~\ref{ch} below~\citep{DOrazio2017}:

\begin{equation}
\chi^2 = \sum_{i=1}^{n} \frac{[mag(t_i)-A\sin \Omega (t_i-t_0) +B]^2}{[\delta mag(t_i)]^2}
\label{ch}
\end{equation}

The Lomb-Scargle (LS) periodogram~\citep{Lomb1976, Scargle1982} is widely used for periodicity detection in unevenly sampled data. In our analysis, we utilized the generalized Lomb-Scargle (GLS) periodogram~\citep{Zechmeister2009} implemented in the \texttt{pyastronomy}~\footnote{https://pyastronomy.readthedocs.io/en/latest/index.html} package. In the GLS periodogram, we calculate a series of ``powers'' defined as Equation~\ref{pow}:

\begin{equation}
p= \frac{\chi_0^2-\chi^2}{\chi_0^2}
\label{pow}
\end{equation}

where $\chi_0^2$ is the residual of a model assuming no variability, and $\chi^2$ is equivalent to Equation~\ref{ch} in the GLS periodogram~\citep{Zechmeister2009}. Compared to the LS periodogram, the GLS periodogram provides a more accurate frequency prediction by taking into consideration an offset and weights, which allows the calculation of the maximum power in the GLS periodogram to be equivalent to minimizing Equation~\ref{ch}. The searched period range of the GLS periodogram is from 700 days to $T_0$ days, where $T_0$ is the time span of the light curve.

\subsection{Selection Criteria}
\label{sc}

We selected reliable periodic candidates based on five criteria. First, to account for stochastic red noise variability, we adopt a comparative approach by measuring the performance of the candidates against that of simulated light curves, as described in Section~\ref{stochastic}. Rather than applying a single threshold to the normalized periodograms, we require $3\sigma$ significance in both the W1 and W2 bands.

Secondly, we adopt the signal-to-noise (S/N) ratio defined as Equation~\ref{cri2} ~\citep{Horne1986}, where $A_0$ denotes the amplitude and $\sigma_r^2$ denotes the variance of the residuals. we require that $\xi>2$ for both the W1 and W2 bands.

\begin{equation}
\xi=A_0^2/(2\sigma_r^2)
\label{cri2}
\end{equation}

Thirdly, we require that the difference between the frequencies of the W1 and W2 light curves must not exceed 5\% of their sum, in accordance with Equation~\ref{cri3}:

\begin{equation}
\frac{\vert f_1 - f_2 \vert}{f_1 + f_2} < 0.05
\label{cri3}
\end{equation}

where $f_1$ and $f_2$ represent the frequencies of the W1 and W2 light curves. In addition, the phase difference between them must be less than 1/8 cycles. This criterion reflects the requirements of the dust-echo model and guarantees that the two sinusoids have the same frequency and phase. However, we allow for a margin of error of 1/8 cycles to account for potential minor phase differences arising from low photometric accuracy or intrinsic characteristics of the IR light curves. For instance, the W2 band may receive a greater contribution from dust with lower temperatures located further from the BH, leading to a slight delay in the W2 variability compared to the W1 band~\citep{Lyu2019}. If the sources are located at high redshifts, i.e. $z>2$, the W1 and W2 emission could mainly come from a relatively inner and outer part of the accretion disk, then a time delay can also be expected. We confirm this small delay for our candidates in Section~\ref{can}.

Fourthly, the available data are required to show more than two cycles in both the W1 and W2 bands. It is important to note that even strictly sinusoidal variations are difficult to distinguish from a simple stochastic process when the number of cycles $N_{\text{cyc}}$ is $\lesssim 5$~\citep{Vaughan2016}. Any finite light curve generated by the corresponding variability process represents only a snapshot of that process. The periodogram derived from this light curve will show power distributed around the actual Power Spectral Density (PSD) continuum with some scatter. This scatter can create the appearance of spikes in the periodogram (e.g., see Fig.~\ref{lc} and Fig.~\ref{lcr}), leading the light curve to seemingly exhibit a sinusoidal-like pattern over 2 to 3 cycles, even if the underlying PSD process is purely continuous. Therefore, these patterns can only be reliably distinguished with much longer time series spanning many (e.g., $>$4--5) cycles of the putative period. However, most optical searches still adhere to a criterion of $>$1.5 cycles due to the limited time span of the data (e.g.~\citealt{Graham2015a, Charisi2016, Liu2019, Chen2024}). Consequently, false positives would arise in the candidate sample from stochastic red noise variability~\citep{Vaughan2016}. Due to the limited number of detected epochs, which amounts to 23, we set the limit to $>$two cycles. We believe that this is the best approach we can take with the current WISE data, but we note that further improvements should be pursued in future research with more available data.

\begin{figure}
\centering
\begin{minipage}{0.5\textwidth}
\centering{\includegraphics[width=1.0\textwidth]{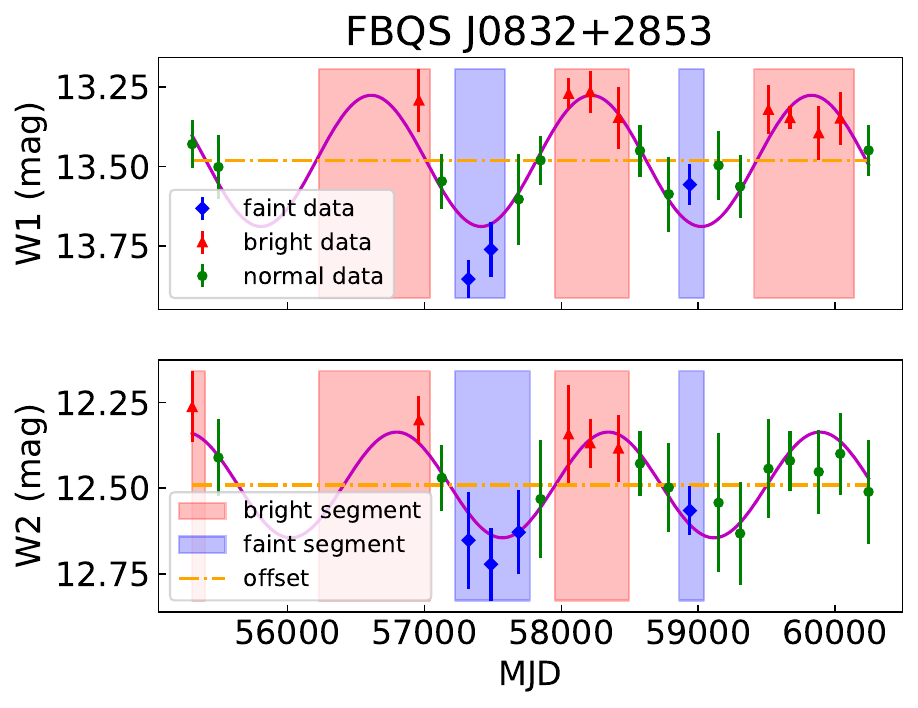}}
\end{minipage}
\caption{An example for selection Criterion 5. In this example, the total count for W1 is 4, while W2 has 3 counts. Note that the errorbars show $2\delta mag(t_i)$, and the purple lines show the best-fit sine curves.}
\label{ex}
\end{figure}

Fifthly, we check if the maximum power of the periodogram indicates a ``correct'' period, since anomalous data points (e.g., an outlier with a small error) could mislead the periodogram calculation. Additionally, this criterion allows us to check if the light curves show strong variability. We estimate the period of a light curve by examining the data points. If all data points within a time segment satisfy $mag(t_i)-2\delta mag(t_i)>$ offset (calculated by the periodogram), we classify it as a ``faint segment''. If all data points within a time segment satisfy $mag(t_i)+2\delta mag(t_i)<$ offset, we classify it as a ``bright segment''. Then we conduct a count in chronological order, recording each change from a faint segment to a bright segment or vice versa, and neglecting normal data points in this process (see an example shown in Fig.~\ref{ex}). Denoting the total count as $n$, ideally $n/2$ periods would be included in the time span. We require that for both the W1 and W2 bands, the period at the maximum power $T_{max}$ and the time span of the light curve $T_0$ satisfy Equation~\ref{cri1}:

\begin{equation}
\frac{max\{n - 2, 0\}}{2} < \frac{T_0}{T_{max}} < \frac{n + 3}{2}
\label{cri1}
\end{equation}

Fig.~\ref{ex} illustrates an AGN that meets all the selection criteria except for the selection Criterion 5 in the W2 band. The periodogram analysis reveals a $3\sigma$ significance in the W2 light curve, indicating a period of $T_{\text{max}} = 1545.6$ days within a time span of $T_0 = 4939.9$ days. However, the total count for W2 is 3, so that it does not meet Criterion 5. This failure is attributed to the lack of significant variability in the W2 band, as shown in Fig.~\ref{ex}.

\section{Analysis and Results}
\label{results}

\subsection{Simulating with Stochastic Variability}
\label{stochastic}

\begin{figure}
\centering
\begin{minipage}{0.5\textwidth}
\centering{\includegraphics[width=1.0\textwidth]{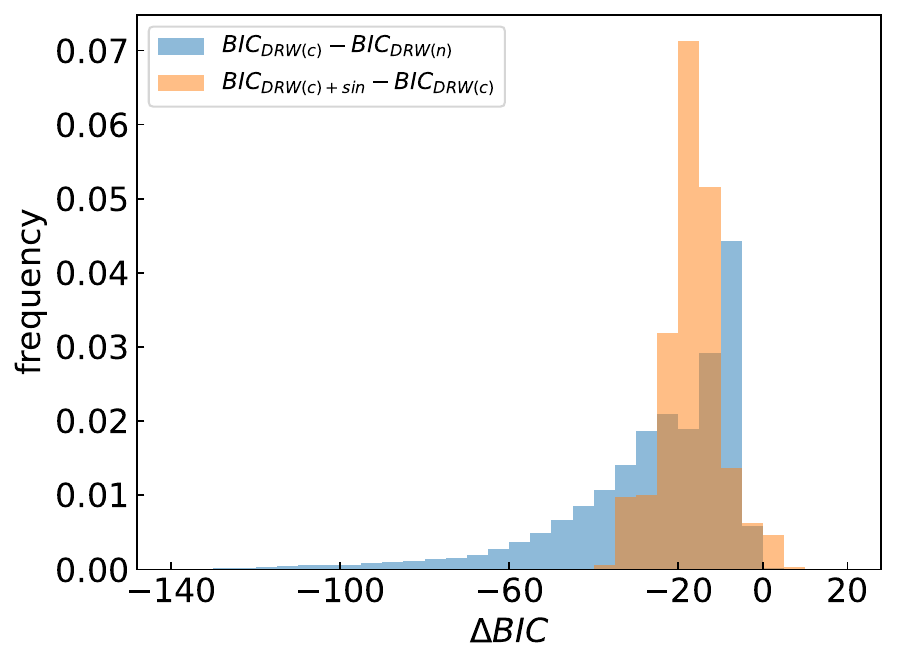}}
\end{minipage}
\caption{Distribution of the $\Delta$ BIC value. $BIC_{\text{DRW}(c)}$ is the BIC value of correlated DRW model, $BIC_{\text{DRW}(n)}$ is the BIC value of the uncorrelated DRW model, while $BIC_{\text{DRW}(c) +sin}$ is the BIC value of the sinusoid +correlated DRW model for the candidates.}
\label{pbic}
\end{figure}

AGN variability can roughly be described by a Gaussian first-order continuous autoregressive model (CAR(1),~\citealt{Kelly2009}), also known as a damped random walk (DRW,~\citealt{MacLeod2010}) or Ornstein-Uhlenbeck process. The likelihood function for the DRW process can be represented as:

\begin{equation}
\mathcal{L} \varpropto \vert C\vert^{-\frac{1}{2}}\exp\left[-\frac{1}{2}\sum_{i, j}(X_i-q)(C^{-1})_{ij}(X_j-q)\right]
\label{lf}
\end{equation}

where the $X_i$ represents the flux of the data point $i$, and $q$ is the long-term mean of the light curve. The covariance matrix $C$ is given by:

\begin{equation}
C_{ij} = \delta_{ij}\sigma_i^2 + \sigma^2\exp\left(-\frac{\left|t_i-t_j\right|}{\tau}\right)
\label{ncf}
\end{equation}

where $\delta_{ij}$ is the Kronecker delta. In our analysis, we assume a strong correlation between the W1 and W2 data. Subsequently, we have devised a modified form of the covariance matrix that accounts for this correlation, expressed as:

\begin{equation}
C_{ai,bj} = \delta_{ab}\delta_{ij}\sigma_i^2 + \left[\rho +(1-\rho)\delta_{ab}\right]\sigma^2\exp\left(-\frac{\left|t_i-t_j\right|}{\tau}\right)
\label{ccf}
\end{equation}

where $a$ and $b$ represent the observed band (W1 or W2) of the data points, and $\rho$ is the correlation coefficient.

\begin{figure*}
\centering
\begin{minipage}{1.0\textwidth}
\centering{\includegraphics[width=0.49\textwidth]{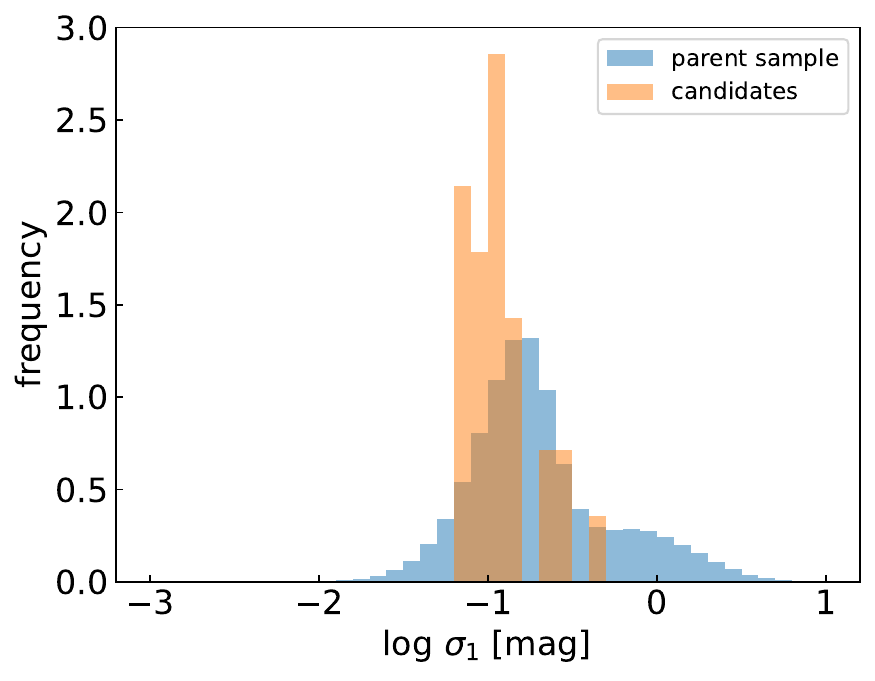}}
\label{si1}
\centering{\includegraphics[width=0.49\textwidth]{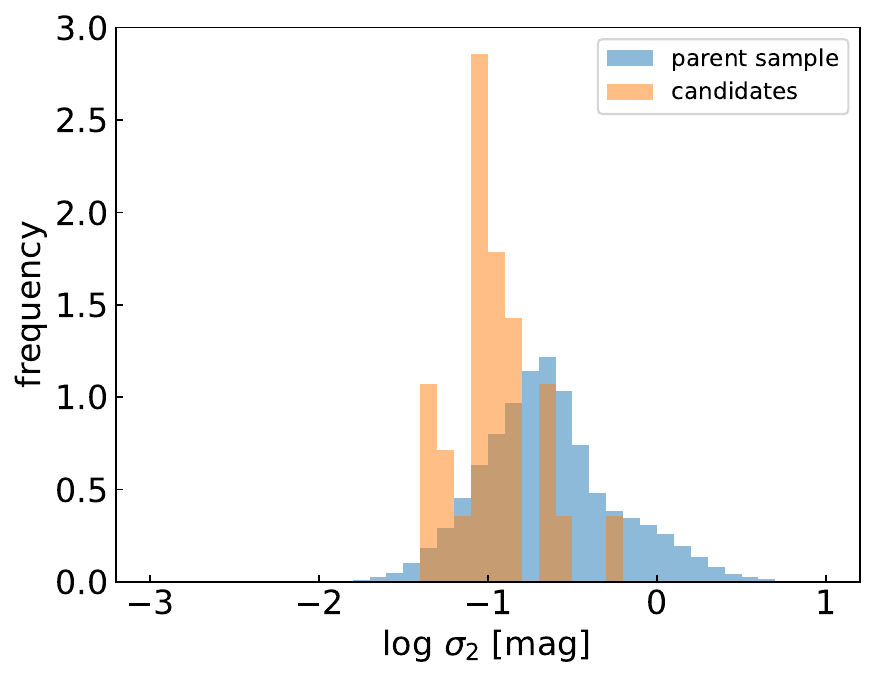}}
\label{si2}
\end{minipage}
\begin{minipage}{1.0\textwidth}
\centering{\includegraphics[width=0.49\textwidth]{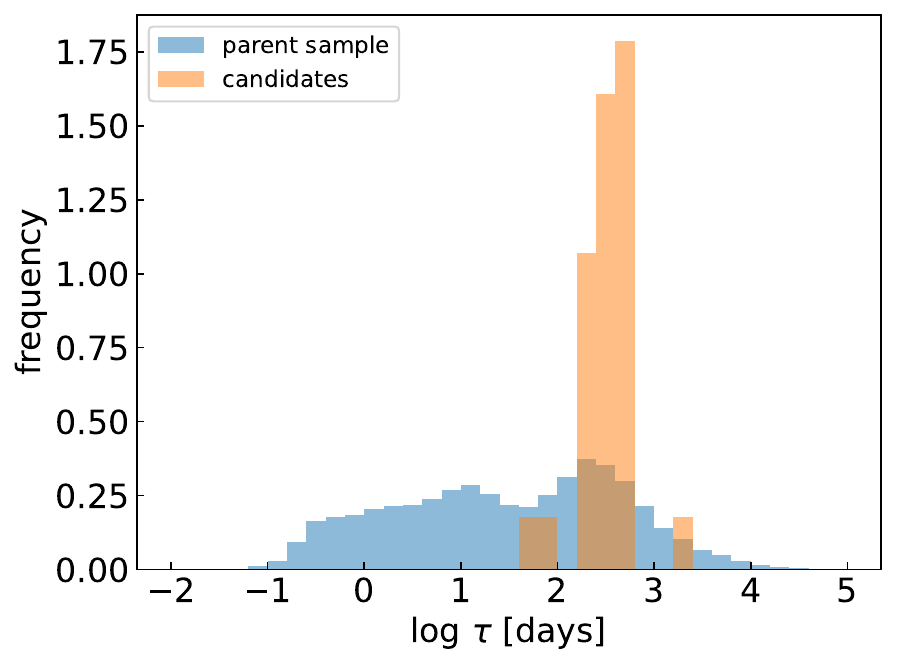}}
\label{ta}
\centering{\includegraphics[width=0.49\textwidth]{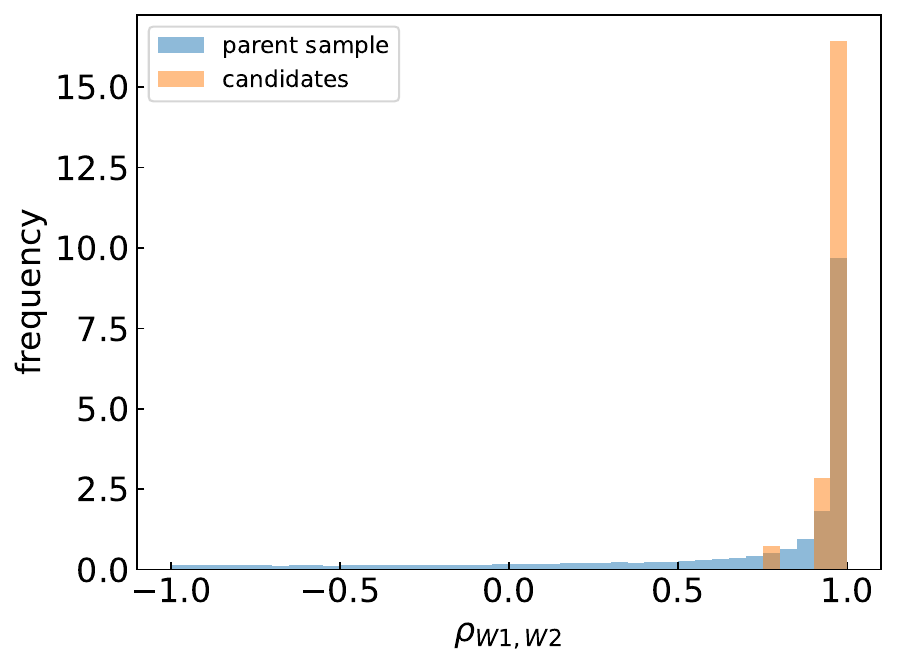}}
\label{rh}
\end{minipage}
\caption{The DRW parameters distribution for the selection parent sample (blue) and the candidates (orange). Top: the $\sigma$ distribution in the W1 band (left) and the W2 band (right). Bottom: the $\tau$ distribution (left) and the correlation coefficient distribution (right).}
\label{pa}
\end{figure*}

Regarding the data quality of WISE, we only use the binned data when estimating DRW parameters. We set a uniform prior for the logarithm of the parameters $\sigma_1$, $\sigma_2$ and $\tau$, and a uniform prior for $\rho_{W1, W2}$. We set $q$ as the mean magnitude of the light curve to reduce computational cost and use the Markov Chain Monte Carlo (MCMC) sampler \texttt{emcee}~\footnote{https://emcee.readthedocs.io/en/stable/}~\citep{Foreman-Mackey2013} to construct the posterior samples of the parameters. However, we find that some light curves lack enough variability but pass the variability criterion (see Section~\ref{sample}) due to certain data points with very small errors. Consequently, they are not adequately constrained by the DRW model. To ensure a stable posterior for the MCMC processes, we set a criterion that the $1\sigma$ error of the posterior $lg\tau$ sample must be less than 2. This is the final requirement to define our parent sample, resulting in a final number of 48,932 (see Section~\ref{sample}). The statistical distributions of $\sigma$, $\tau$, and $\rho_{W1, W2}$ are illustrated in Fig.~\ref{pa}. Upon comparison, the selected candidates exhibit similar values for $\sigma_1$, $\sigma_2$, and $\tau$ compared to the parent sample. Notably, they display a stronger correlation, which aids in meeting selection criterion 4 easier. The distribution of $\rho_{W1, W2}$ serves as validation for our correlation assumption.

As an alternative test, we perform a model selection to test if the correlation parameter is necessary to explain the light curve on top of a uncorrelated stochastic background. We adopt a maximum likelihood approach for the model comparison and parameter estimation. We use the
Bayesian information criterion (BIC), which is defined as:

\begin{equation}
BIC = -2\ln\mathcal{L} +k\ln N
\label{bic}
\end{equation}

where $\mathcal{L}$ is the likelihood function, $k$ is the number of free model parameters, and $N$ is the number of data points. A lower BIC value indicates a preferable model. Typically, when $\Delta BIC < -10$, the model with the lower BIC is strongly favored. For the uncorrelated DRW process, we generate posterior parameter samples independently for each band. Subsequently, we integrate these samples into a covariance matrix to compute $\mathcal{L}$ with all correlation components set as zero. Our analysis indicates a strong preference for the correlated model over the uncorrelated one, aligning with our earlier analysis (further details provided in Fig.~\ref{pbic}).

We proceed by randomly selecting a set of parameters to generate two correlated W1 and W2 simulated light curves. These simulated light curves are then compared with the actual observation cadence and measurement errors. This process is repeated 100,000 times to check that the selection criteria are met. In addition, we use the periodogram values of these simulated light curves to determine the significance of our candidates, as described in Criterion 1 in section~\ref{sc}. To further validate our simulation methodology, we perform a statistical comparison between the real and simulated light curves. Our analysis shows that they exhibit similar behaviour for each selection criterion (detailed results in Table~\ref{st}), suggesting that the correlated DRW processes could potentially generate a comparable number of candidates to the real ones.

\begin{table}
\centering
\caption{Statistical comparison between real light curves and simulated ones.}
\begin{tabular}{ccc} \hline
Constraints & Real & Mock \\ \hline
Numbers of AGNs & 48,932 & 48,932 \\
W1, 3$\sigma$ significance & 1,075 & 1,360 \\
W2, 3$\sigma$ significance & 1,099 & 1,483 \\
W1, $\xi>2$ & 6,360 & 8,580 \\
W2, $\xi>2$ & 6,724 & 9,023 \\
Close frequency \& phase & 19,590 & 21,194 \\
W1, Show >2 cycles & 22,107 & 23,192 \\
W2, Show >2 cycles & 21,987 & 23,561 \\
W1, correct period & 28,345 & 25,340 \\
W2, correct period & 27,886 & 25,145 \\ \hline
candidates & 28 & 51.22 \\ \hline
\end{tabular}
\tablecomments{Data in the ``Mock'' column refer to 1 /100,000 of the total from 4,893,200,000 simulated light curves, allowing direct comparison with the real data. ``3$\sigma$ significance'' refers to the selection criterion 1 in Section~\ref{sc}. ``correct period'' refers to the selection criterion 5 in Section~\ref{sc}.}
\label{st}
\end{table}

Notably, the DRW model is relatively simple and may not fully capture the complexity of actual AGN variability due to degeneracies~\citep{Zu2013}. When using the DRW assumption to calculate false alarm probabilities, it might not completely consider the assumptions of the red noise hypothesis. However, using different variability models could lead to some small changes in the results, but usually not significant ones.

\subsection{Candidates of Periodic Sources}
\label{can}

\begin{figure}
\centering
\begin{minipage}{0.5\textwidth}
\centering{\includegraphics[width=1.0\textwidth]{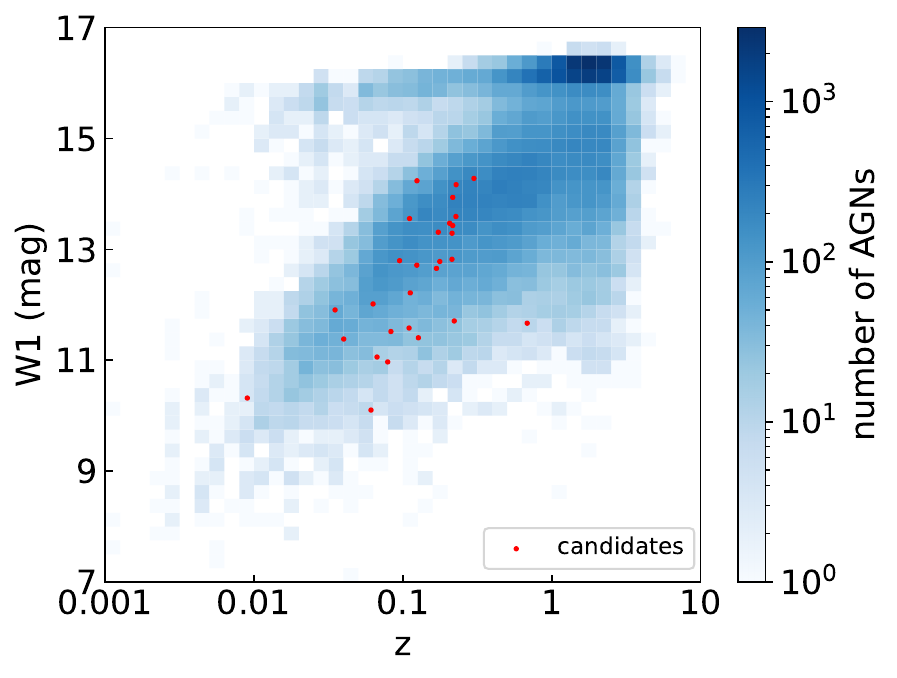}}
\end{minipage}
\caption{Distribution of the parent sample (blue), and the candidates (red) in the redshift and W1 magnitude space. We use a $40\times 40$ grid to visualize the distribution of the parent sample. The color for each grid cell represents the number of AGNs from the parent sample that fall within that cell.}
\label{z}
\end{figure}

Using the data and criteria outlined in Section~\ref{data}, we have identified a total of 28 candidates. The catalog of candidates can be found in Table~\ref{cat}. The light curves and the GLS periodograms of two examples are presented in Fig.~\ref{lc} and the rest can be found in Fig.~\ref{lcr}. The W1 and redshift distributions of both the parent sample and the selected candidates are shown in Fig~\ref{z}.

It is worth noting that the candidates are predominantly situated in the low redshift range (median z=0.126 , with 27/28 at $z<$0.4) and bright region (median W1=12.69, 25/28 satisfy W1$<$14), which can be attributed to a selection effect making it easier for them to fulfill the selection criteria.

\begin{figure}
\centering
\begin{minipage}{0.5\textwidth}
\centering{\includegraphics[width=1.0\textwidth]{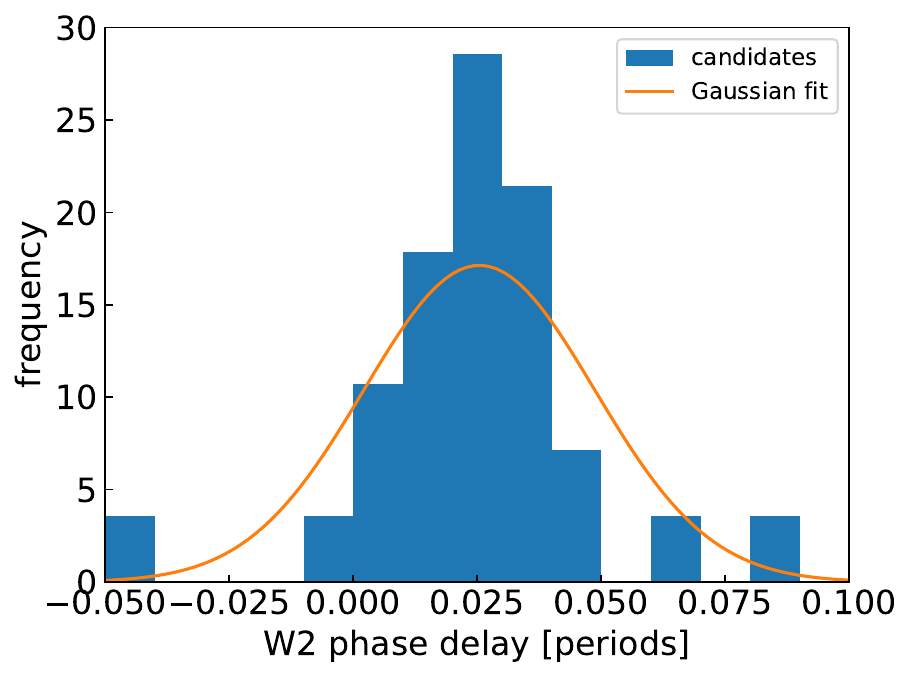}}
\end{minipage}
\caption{The distribution of the phase delay of W2 relative to W1 for the candidates (blue), and the corresponding Gaussian fit of the distribution (orange line).}
\label{pd}
\end{figure}

In Section~\ref{sc}, we mention the possibility of a slight phase delay between the W2 and W1 light curves. This is confirmed by our candidates, with an average delay of 0.025 periods, as illustrated in Fig.~\ref{pd}. The maximum phase delay is 0.087 period, which remains smaller than the 1/8 period threshold used in our selection process, indicating a reasonable choice.

As an alternative test, here we use the BIC method (see details in Section~\ref{stochastic}) to assess if an additional periodic signal is needed to explain the light curve on top of a stochastic background. The likelihood function for the DRW + sinusoidal model is written as:

\begin{equation}
\mathcal{L} \varpropto \vert C\vert^{-\frac{1}{2}}\exp\left[-\frac{1}{2}\sum_{i, j}(X_i-S_i)(C^{-1})_{ij}(X_j-S_j)\right]
\label{lfs}
\end{equation}

where $S_i$ represents the sinusoid signal. Our analysis shows that, for the vast majority of the candidates, $BIC_{\text{DRW}+\text{sin}} - BIC_{\text{DRW}(c)} <-10$, indicating robust evidence that the periodic model is significantly favored over the pure stochastic model (refer to Fig.~\ref{pbic} for detailed results). However, this could be explained by the limitation of the DRW model, which might not adequately capture a spiky PSD and may be more suited to smooth continuum-like PSD shapes. Therefore, additional data in the future are necessary to obtain a higher density in frequency sampling for the periodogram and better test the preference for the sinusoid + DRW model.

It's worth noting that the sinusoid +DRW model is a commonly used model due to its simplicity and clear physical picture. However, there exist alternative light curve models, such as the bursty model predicted by circumbinary accretion disk simulations~\citep{Farris2014}. Despite the potential of this bursty model, we did not simulate it for our candidates as we could not observe ``bursty'' characteristics in the WISE light curves.

Here we compile the derived quantities for our analysis, which are presented in Table~\ref{cat}. The table includes essential information such as the Milliquas Name, right ascension (R.A.), declination (Decl.), redshift ($z$), and quasar properties as provided by~\citet{Wu2022} and \citet{Liu2019apjs}. Additionally, we have included periodic properties such as offsets of the W1 and W2 bands, and the period of the candidates.

\subsection{False Alarm Probability}
\label{fap}

To estimate the false alarm probability (FAP) for each candidate, we employ the method described in~\citet{Chen2020}. We estimate the approximate FAP using the effective number of independent frequencies $N_{eff}$, which is calculated by dividing the observed frequency window by the expected peak width $\delta f= 1/T$. The FAP is estimated as (e.g.,~\citealt{VanderPlas2018}):

\begin{equation}
FAP \sim 1 - [P_{single}]^{N_{eff}}
\label{fap1}
\end{equation}

Where $P_{single}=1-e^Z$, $Z$ is the periodogram value defined in~\citet{Zechmeister2009}. Since candidates are selected based on the light curves of both the W1 band and W2 band, there are two $P_{\text{single}}$ and $N_{\text{eff}}$. To simplify the calculation, we can estimate the periodogram FAP as Equation~\ref{fap2}:

\begin{equation}
FAP \sim 1 - [P_{single, W1}]^{N_{eff, W1}} [P_{single, W2}]^{N_{eff, W2}}
\label{fap2}
\end{equation}

It is important to acknowledge that this calculation assumes independent FAPs for the W1 and W2 bands, contrary to the strong correlation observed in the DRW simulation (refer to Section~\ref{stochastic}). Nevertheless, this method provides an upper limit estimate for the FAP, as single-band FAP calculations do not consider this correlation.

\startlongtable
\begin{deluxetable*}{ccccccccccc}
\tablecaption{The 28 periodic AGN candidates meeting the selection criteria.}
\label{cat}
\tablehead{
\colhead{Milliquas Name} & \colhead{R.A.} & \colhead{Decl.}  & \colhead{z} & \colhead{W1} & \colhead{W2} &  \colhead{period} & \colhead{Type} & \colhead{log \lbol} & \colhead{log($\frac{M_{BH}}{M_{\odot}}$)} & \colhead{log \lamedd}
}
\colnumbers

\startdata
PGC 3095405 & 4.3447 & 5.3532 & 0.11 & 11.582 & 10.585 & 1811 & AX & - & - & - \\
PGC 1523911 & 10.1054 & 17.0707 & 0.112 & 12.216 & 11.578 & 1811 & AR & - & - & - \\
ZOAG 139.33-5.77 & 41.0125 & 53.4745 & 0.035 & 11.909 & 11.016 & 2186 & NX & - & - & - \\
1WGA J0536.2+6027 & 84.0463 & 60.4566 & 0.684 & 11.67 & 11.133 & 1428 & QRX & - & - & - \\
ESO 254-16 & 91.6785 & -43.7311 & 0.04 & 11.383 & 10.916 & 1268 & NR & - & - & - \\
2MASS J06263670-4258059 & 96.6529 & -42.9683 & 0.3 & 14.281 & 13.635 & 2019 & BRX & - & - & - \\
RXS J07570+5832 & 119.2705 & 58.5445 & 0.168 & 12.658 & 11.709 & 1865 & QX & - & - & - \\
SDSS J101241.21+215556.1 & 153.1717 & 21.9322 & 0.111 & 13.556 & 12.661 & 1863 & A & 44.46 & 7.79 & -1.445 \\
2MASX J11005099+5135026 & 165.2126 & 51.584 & 0.214 & 13.292 & 12.472 & 1829 & ARX & 45.19 & 8.69 & -1.614 \\
ESO 377-24 & 168.139 & -36.4253 & 0.009 & 10.319 & 9.8026 & 1592 & AR & - & - & - \\
ChaMP J115907.1+290842 & 179.7798 & 29.1452 & 0.227 & 13.594 & 12.66 & 2274 & NRX & - & - & - \\
SDSS J124731.75+450121.3 & 191.8823 & 45.0226 & 0.214 & 12.822 & 11.937 & 2019 & ARX & 45.16 & 8.55 & -1.498 \\
SDSS J140336.43+174136.1 & 210.9018 & 17.6934 & 0.221 & 11.71 & 10.797 & 2346 & QRX & 45.81 & 9.63 & -1.931 \\
SDSS J143640.16+422933.8 & 219.1674 & 42.4927 & 0.216 & 13.939 & 13.061 & 2319 & AR & 44.9 & 8.46 & -1.673 \\
DESI 39632956414755490 & 219.6496 & 33.6835 & 0.216 & 13.43 & 12.487 & 2019 & Q & - & - & - \\
SDSS J162938.09+362452.2 & 247.4087 & 36.4145 & 0.124 & 14.238 & 13.433 & 1829 & AR & 44.28 & 8.5 & -2.338 \\
SDSS J165822.32+183735.1 & 254.593 & 18.6264 & 0.177 & 12.782 & 11.957 & 2229 & ARX & 45.28 & 8.38 & -1.213 \\
IRAS 17020+4544 & 255.8766 & 45.6798 & 0.061 & 10.102 & 9.1251 & 1706 & ARX & - & - & - \\
SDSS J171640.99+270544.1 & 259.1708 & 27.0956 & 0.228 & 14.168 & 13.223 & 2437 & AX & 44.34 & 6.8 & -0.573 \\
TXS 1919-166 & 290.6449 & -16.5482 & 0.127 & 11.406 & 10.387 & 1941 & AR2 & - & - & - \\
6dF J193819.6-432646 & 294.5817 & -43.4462 & 0.079 & 10.971 & 10.01 & 1903 & AR & - & - & - \\
6dF J202557.4-482226 & 306.4891 & -48.3739 & 0.067 & 11.059 & 10.197 & 2251 & A & - & - & - \\
6dF J203052.8-345259 & 307.7199 & -34.883 & 0.124 & 12.716 & 11.78 & 2186 & AR & - & - & - \\
IGR J20450+7530 & 311.1435 & 75.533 & 0.095 & 12.797 & 12.084 & 2251 & AX & - & - & - \\
SDSS J211655.57-013435.4 & 319.2316 & -1.5765 & 0.206 & 13.472 & 12.557 & 1268 & A & 44.91 & 8.36 & -1.562 \\
6dF J214907.4-175159 & 327.2808 & -17.8665 & 0.063 & 12.017 & 11.144 & 2039 & AX & - & - & - \\
SDSS J215644.32-083529.3 & 329.1847 & -8.5915 & 0.173 & 13.311 & 12.323 & 2297 & A & 44.83 & 7.92 & -1.207 \\
PGC 3096701 & 342.4148 & 11.008 & 0.083 & 11.519 & 10.668 & 1758 & AX & - & - & - \\
\enddata
\tablecomments{
(1)-(4): Name, RA, DEC, and redshift of the objects.
(5)-(6): Median magnitudes in the W1 and W2 bands, respectively.
(7): Period (in the observer-frame) in units of days.
(8): Legend of type/class from~\citet{Flesch2023}:
Q = QSO, type-I broad-line core-dominated, 860,100 of these.
A = AGN, type-I Seyferts/host-dominated, 47,044 of these.
B = BL Lac type object, 2,814 of these. (FSRQs are typed as QSOs here)
K = NLQSO, type-II narrow-line core-dominated, 6,048 of these.
N = NLAGN, type-II Seyferts/host-dominated, 39,768 of these. Incomplete, and includes an unquantified residue of legacy NELGs/ELGs/LINERs, plus some unclear AGN. This is the catch-all category.
S = star classified but showing quasar-like photometry and radio/X-ray association, thus included as a quasar candidate; 124 of these.
R = radio association displayed.
X = X-ray association displayed.
2 = double radio lobes displayed (declared by data-driven algorithm).
(9)-(11): Logarithmic bolometric luminosity in units of erg/s, black hole mass in units of solar mass, and Eddington ratio for sources in the SDSS DR16 quasar catalog~\citep{Wu2022}, the DR7 board-line AGN catalog~\citep{Liu2019apjs}.
}
\end{deluxetable*}

\begin{figure}
\centering
\begin{minipage}{0.5\textwidth}
\centering{\includegraphics[width=1.0\textwidth]{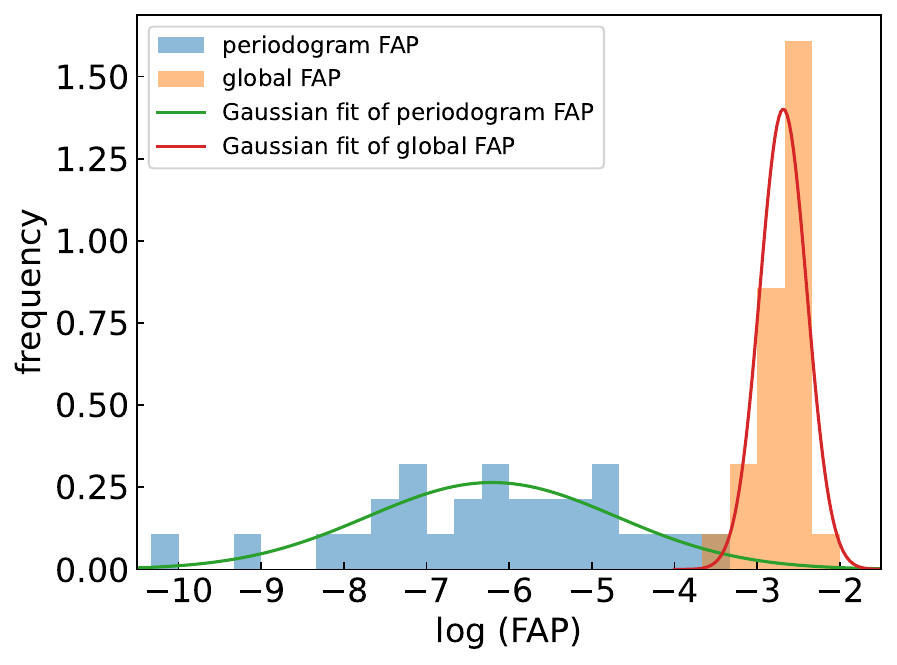}}
\end{minipage}
\caption{The FAP distribution estimated from periodogram calculation (blue), and the global FAP estimated from 100,000 simulated light curves (orange). The Gaussian fit curves for the 2 distributions are also shown.}
\label{ff}
\end{figure}

Additionally, we empirically compute the global FAP using the 100,000 simulated light curves for each AGN, following a similar approach as presented by~\citet{Barth2018}. The global FAP accounts for all false positives that satisfy all selection criteria within the explored frequency range. The global FAP values for candidates range from $4.0 \times 10^{-4}$ to $4.77 \times 10^{-3}$, with a mean of $2.07 \times 10^{-3}$ by Gaussian fit. Further details are available in Fig.~\ref{ff}.

It is noteworthy that the global FAPs are significantly higher than the periodogram FAPs. This discrepancy can be attributed to the stronger variability exhibited by the simulated light curves compared to real observations, leading to an increase in the global FAP. Moreover, while most false positives exhibit only a $3\sigma$ significance level, certain candidates display notably stronger significance, such as SDSS J162938.09+362452.2 and 6dF J214907.4-175159 as depicted in Fig.~\ref{lcr}.

\section{Discussion}
\label{discuss}

\subsection{Do the Candidates Show Distinctive Properties?}
\label{quasarcheck}

\begin{figure}
\centering
\begin{minipage}{0.5\textwidth}
\centering{\includegraphics[width=1.0\textwidth]{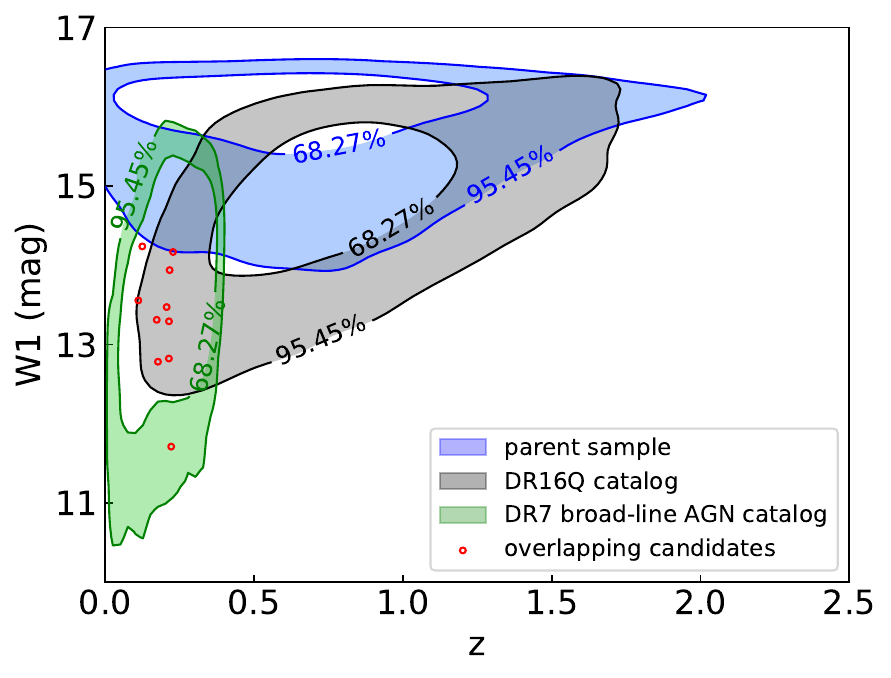}}
\end{minipage}
\caption{Distributions of the selection parent sample (blue), quasars present in the DR16Q catalog that overlap with the parent sample (black), AGNs present in the DR7 broad-line AGN catalog that overlap with the parent sample (green), and the overlapping candidates (red) in the redshift and W1 magnitude space. Enclosed percentiles are marked for each contour level.}
\label{ca}
\end{figure}

One might wonder whether our selected periodic sample shows distinct features in other physical properties, such as bolometric luminosity (\lbol), black hole mass (\mbh) and Eddington ratio (\lamedd=\eddratio). To investigate this, we try to extract a subsample of candidates with previous measurements in these parameters. We chose to cross-match the candidates with two AGN samples based on Sloan Digital Sky Survey (SDSS), namely the SDSS DR16 quasar catalog (DR16Q)~\citep{Wu2022} and DR7 broad-line AGN catalog~\citep{Liu2019apjs}. They contained 750,414 and 14,584 sources, concentrated at high and low redshifts, respectively (see Figure~\ref{ca}). We only found a total of 10 candidates that overlapped with the two catalogs.

\begin{figure*}
\begin{minipage}{1.0\textwidth}
\centering{\includegraphics[width=0.49\textwidth]{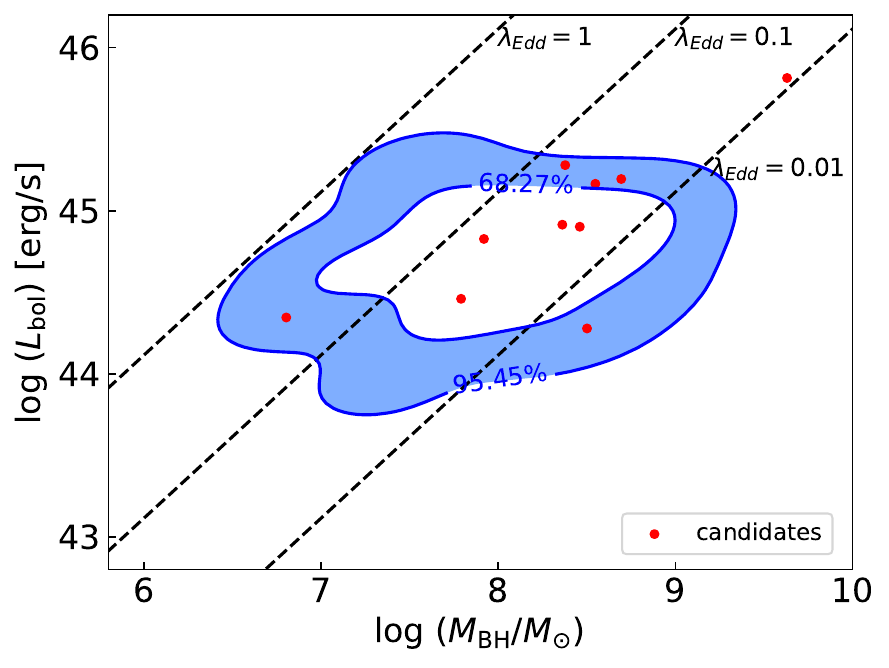}}
\label{dclz}
\centering{\includegraphics[width=0.49\textwidth]{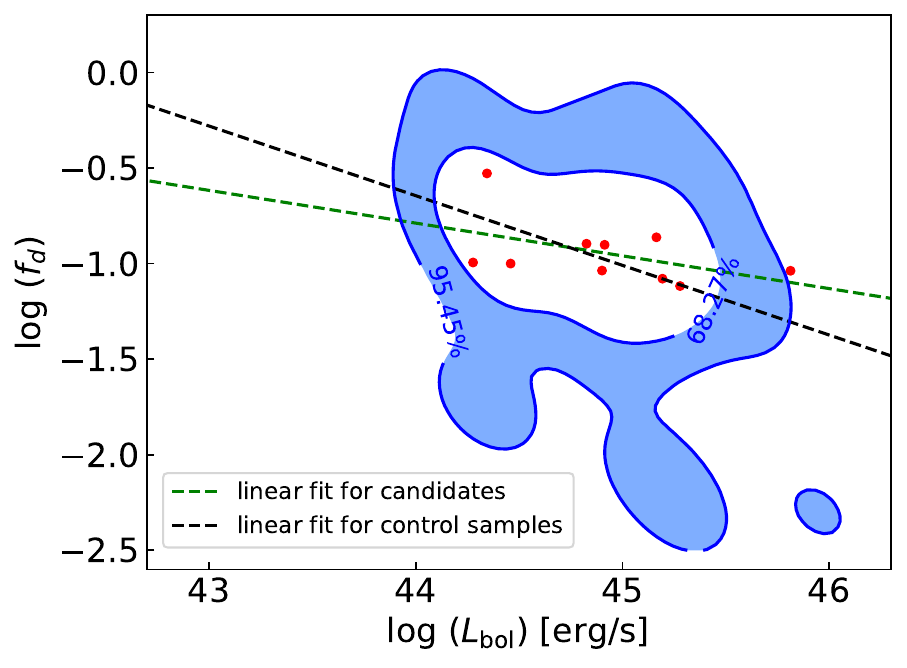}}
\label{dcll}
\end{minipage}
\caption{Left: distribution of the control samples (blue), which control for redshift and W1, W2, and the candidates (red) in the log (\mbh) and log (\lbol) space. Right: distribution of the control samples (blue), which control for redshift and \lbol, and the candidates (red) in the log (\lbol) and log ($f_d$) space (right).The right panel also tests the anti-correlation between \lbol\ and $f_d$.}
\label{pro}
\end{figure*}

Similar to the occupation of the candidates in the whole parent sample (see Figure~\ref{z}), they are all in the low redshift and low magnitude space (see Figure~\ref{ca}) due to selection effects. Therefore, we need to construct a control sample for a fair comparison, at least in redshift and IR magnitudes. Specifically, we selected the 10 closest AGNs with a redshift ratio less than 1.01 and with differences in W1 and W2 magnitudes less than 0.1. In cases when no AGN meet the specified criteria, we choose the closest AGN in the catalog as the control sample. After controlling, we then compare the \mbh, \lbol\ and \lamedd\ of the periodic sources with normal AGNs. Additionally, we have also considered another parameter dust covering factor ($f_d$), defined as the ratio of the W1 monochromatic luminosity ($L_{W1} =\nu_{W1} F_{\nu W1}$) and the \lbol\ ($f_d=\fc$, e.g.,~\citealt{Ma2013}), to indicate the extent of dust obscuration. However, no significant differences in the distributions of these parameters can be seen visually (see Fig.~\ref{pro}). We have also checked the mean and standard deviation (SD) of their distributions (see Table~\ref{ms}) and confirmed it. 

\begin{table}
\centering
\caption{Statistical comparison of quasar properties between candidates and control samples.}
\begin{tabular}{cccc} \hline
& candidates & control 1 & control 2 \\ \hline
mean of log(\lbol) & 44.92 & 44.70 & 44.84 \\
SD of log(\lbol) & 0.45 & 0.29 & 0.37 \\
mean of log(\mbh) & 8.310 & 7.989 & 7.945 \\
SD of log(\mbh) & 0.686 & 0.523 & 0.549 \\
mean of log(\lamedd) & -1.505 & -1.405 & -1.216 \\
mean of log($f_d$) & -0.946 & -0.813 & -0.952 \\
SD of log($f_d$) & 0.160 & 0.224 & 0.399 \\ \hline
\end{tabular}
\label{ms}
\tablecomments{
'control 1' refers to the control sample that controls z, W1 and W2 magnitude. 'control 2' refers to the control sample that controls z and \lbol.
}
\end{table}

We further performed the K-S statistics for the \lbol, \mbh, \lamedd, and $f_d$ distributions between the candidates and the control sample, yielding values of 0.375, 0.400, 0.225, and 0.400, respectively. The K-S tests suggest significant differences in these properties with a critical value of 0.480 for a 95\% confidence level, so none of them show significant differences. It is worth emphasizing that the systematic uncertainties in the estimates of AGN properties (e.g., single-epoch BH mass, see discussions in~\citealt{Shen2013}) are nonnegligible and would dilute the difference, if there were any. This may be particularly important for the small sample size for comparison here.

\subsection{Comparison with SMBHB Candidates Selected by Optical Periodicity}
\label{pre}

As mentioned in the introduction, previous studies have identified samples of SMBHB candidates based on optical periodic light curves. It is intriguing to investigate whether there are any common sources between our IR-selected sample and these optically-selected samples. Such overlap would provide compelling evidence for the periodicity of the candidates. Additionally, if the periods derived from optical light curves align with those obtained from IR light curves, we can employ the dust-echo model to further explore the properties of the candidates. For this purpose, we compile a collection of optical SMBHB candidates reported in various literature, including~\citet{Graham2015a, Charisi2016, Zheng2016, Liu2019, Chen2020, Chen2024}.

\begin{figure}
\centering
\begin{minipage}{0.5\textwidth}
\centering{\includegraphics[width=1.0\textwidth]{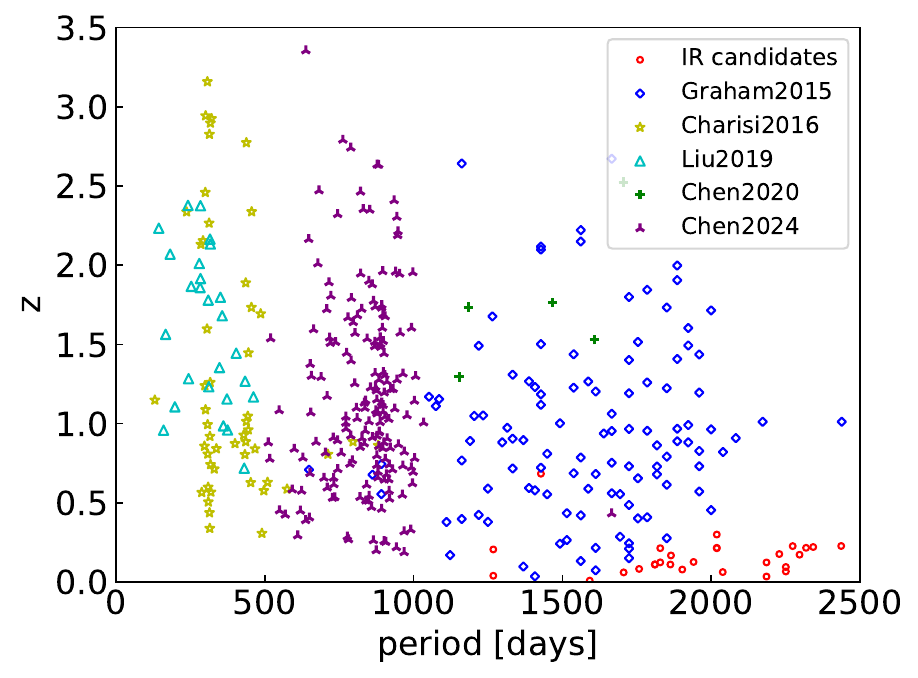}}
\end{minipage}
\caption{Redshift and period distributions of SMBHB candidates selected by this work (red), and those reported in previous works by optical search (other colors).}
\label{candi}
\end{figure}

Unfortunately, we do not find any overlap between our sample and those samples. The lack of matches is not entirely unexpected, as the two searches prioritize candidates with different periods. Specifically, we find that the periods of the optical candidates predominantly fell below 1700 days, whereas the periods of our selected candidates were mostly longer than 1700 days (see details shown in Fig.~\ref{candi}). Additionally, most of our candidates have low redshifts (z$<$0.4) due to the detection ability of WISE, while the redshifts of candidates from optical searches span from 0 to $\sim$3.

\subsection{Optical Light Curves and A Possible Periodic Source}

\begin{figure}
\centering
\begin{minipage}{0.5\textwidth}
\centering{\includegraphics[width=1.0\textwidth]{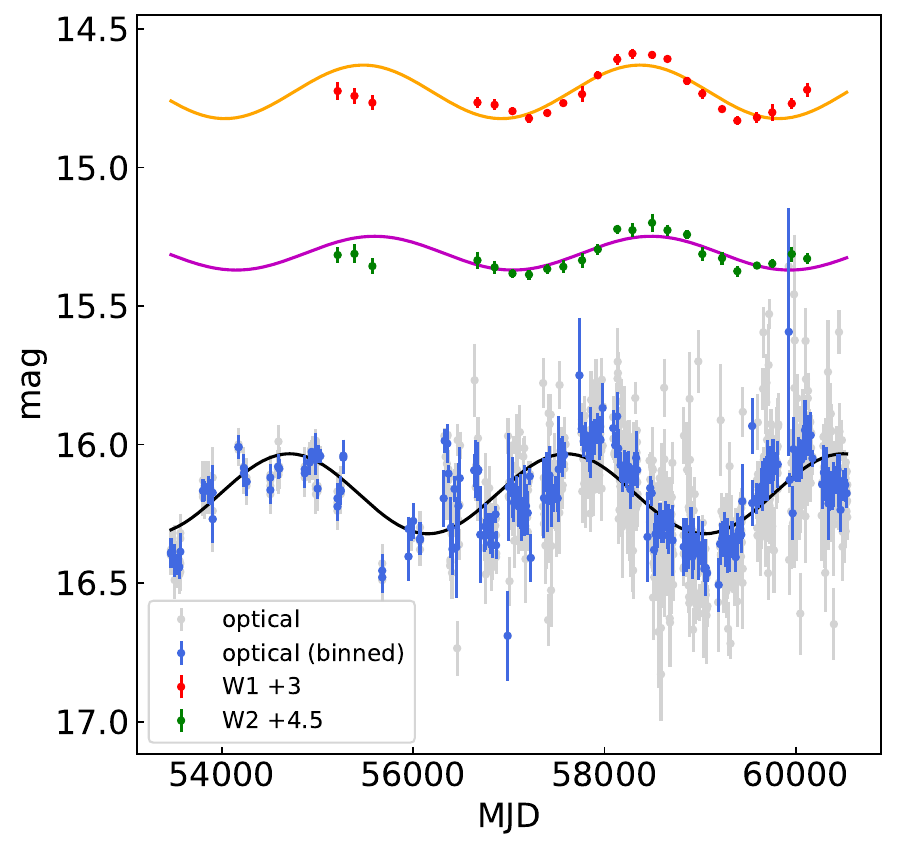}}
\end{minipage}
\caption{The W1, W2 and optical light curves of SDSS J140336.43+174136.1. The orange, purple and black lines represent the corresponding best-fitting sinusoids. We assume that the periods of IR light curves are the same as that of the optical light curves.}
\label{op}
\end{figure}

Given the abundance of optical photometric archival data, we then examined the optical light curves of the 28 candidates using data from various surveys, such as the Catalina Real Time Transient Survey (CRTS, ~\citealt{Drake2009}), the Palomar Transient Factory (PTF,~\citealt{Rau2009}), the Zwicky Transient Facility (ZTF,~\citealt{Masci2019}), and the All-Sky Automated Survey for Supernovae (ASAS-SN,~\citealt{Kochanek2017}). Their combined dataset provides optical light curves spanning about two decades. We have normalized the data from different surveys and binned them at 20-day intervals. However, they generally show a small amplitude of optical variability, probably due to either intrinsically weak variability in the optical band or too shallow surveys. Only one candidate, SDSS~J140336.43+174136.1, shows significant variability and periodicity in its optical light curve (see Fig.~\ref{op}).

The periodogram analysis of the optical light curve of SDSS~J140336.43+174136.1 reveals a period of $2891\pm 22$ days, that is slightly longer than the period fitted from the IR light curves, which is $2346\pm 87$ days. The errors are given by periodogram peak statistics based on 100,000 perturbed light curves for each real light curve. The discrepancy could be caused by the poor cadence of the WISE surveys, and so in the following analysis we choose the optical period for this particular source.

We present the corresponding best-fitting sinusoids for the W1, W2, and optical light curves in Figure~\ref{op}. The time delay between the W1 (W2) and optical light curves is 630 (727) days (in the rest frame), respectively. Assuming the simplest scenario of a hollow spherical shell of dust with the source located at its center, we can easily estimate that the inner radius of the shell is 0.53 pc. These values are roughly consistent with previous statistical correlations between the time delay and bolometric luminosity of AGNs. For example, using the relations obtained by~\citet{Lyu2019}, i.e. 

\begin{equation}
\Delta t_{W1} /\text{day} =10^{2.10\pm 0.06} (\lbol /10^{11} L_{\odot})^{0.47\pm 0.06}
\label{delay1}
\end{equation}

for the W1 band and

\begin{equation}
\Delta t_{W2} /\text{day} =10^{2.20\pm 0.06} (\lbol /10^{11} L_{\odot})^{0.45\pm 0.05}
\label{delay2}
\end{equation}

for the W2 band, and a given bolometric luminosity of $10^{45.81}$~\lum\ for SDSS~J140336.43+174136.1 (see Table~\ref{cat}), the W1 and W2 time delays are estimated to be $476\pm 147$ days and $566\pm 158$ days, respectively, which is in good agreement with our measurements.

\section{Conclusion}
\label{conclusion}

In this work, we have conducted the first systematic search for SMBHBs in the IR band and identified 28 AGNs exhibiting IR periodic variability using the decade-long \wise\ and \neowise\ light curves. We performed extensive simulations to evaluate the potential influence of stochastic variability on candidate selection. By counting all the false positives in the simulated light curves, the probability of these periodic light curves being generated by random processes is an average of 0.207\% by Gaussian fit. However, the number can be reproduced by our mock simulations with the DRW process of the parent sample. This indicates the challenge of identifying reliable SMBHBs with periodic emission using only WISE light curves, which are constrained by their time span and visit cadence, necessitating a
low threshold on the minimum number of cycles ($N_{\text{cyc}}>2$). Consequently, the nature of these periodic sources, whether driven by SMBHBs or not, should be carefully tested in future observations. We subsequently investigated whether these IR periodic sources exhibit distinct properties compared to normal AGNs. However, no significant biases were found, and the small sample size prevents us from statistically identifying potential weak differences. On the other hand, we found that there is no overlap between our sample (IR-selected) and the SMBHB candidates reported in the literature (optically selected), likely due to differences in their preferred period ranges.  Interestingly, we have identified SDSS J140336.43+174136.1 as a candidate displaying periodicity in both optical and IR bands. Further extended and more sensitive observations of these candidates are essential to confirm whether or not they are real periodic sources in both optical and IR bands. As suggested by~\citet{DOrazio2017}, the periodic sources can also help test the physical mechanisms that produce the observed periodicity, due to relativistic Doppler modulation or accretion rate fluctuations.

This study highlights the promising potential of identifying SMBHB candidates through IR time-domain observations, which can significantly complement searches in other bands, although there are challenges in distinguishing genuine sources. It opens up a new avenue for SMBHB searches and can be applied to future surveys. Although \neowise\ unfortunately ended on July 31, 2024, the upcoming NEO Surveyor~\citep{NEOsurveyor2023}, the successor to \neowise, and the Nancy Grace Roman Space Telescope~\citep{Roman2015} will ensure a bright future for IR time-domain astronomy, particularly for SMBHB searches~\citep{Haiman2023}. These different surveys somewhat form a relay in time and provide us with IR light curves with a long time baseline, which is essential for verifying the periodicity of the candidates with $N_{\text{cyc}}<4$ and for detailed comparisons against physical models.

\begin{acknowledgments}

We sincerely thank the expert referee for very constructive and insightful comments, which helped improve our manuscript greatly.
This work is supported by the Strategic Priority Research Program of the Chinese Academy of Sciences (Grant No. XDB0550200), the National Natural Science Foundation of China (grants 12073025, 12192221, the science research grants from the China Manned Space Project and the Cyrus Chun Ying Tang Foundations. D.L. acknowledges the support from the National Undergraduate Training Program for Innovation and Entrepreneurship. X.L. acknowledges support by NSF grant AST-2206499. This research makes use of data products from the Wide-field Infrared Survey Explorer, which is a joint project of the University of California, Los Angeles, and the Jet Propulsion Laboratory/California Institute of Technology, funded by the National Aeronautics and Space Administration. This research also makes use of data products from NEOWISE-R, which is a project of the Jet Propulsion Laboratory/California Institute of Technology, funded by the Planetary Science Division of the National Aeronautics and Space Administration. This research has made use of the NASA/IPAC Infrared Science Archive, which is operated by the California Institute of Technology, under contract with the National Aeronautics and Space Administration. 

\end{acknowledgments}

\bibliographystyle{mnras}
\bibliography{ref}

\appendix
\begin{appendices}
\twocolumngrid

\section{The IR Light Curves of the Periodic AGN Sample}

We have shown the light curves of 2 special candidates in Fig.~\ref{lc}. Here we show the light curves of the rest 26 candidates in Fig.~\ref{lcr}.

\begin{figure*}
\centering
\begin{minipage}{1.0\textwidth}
\centering{\includegraphics[width=0.49\textwidth]{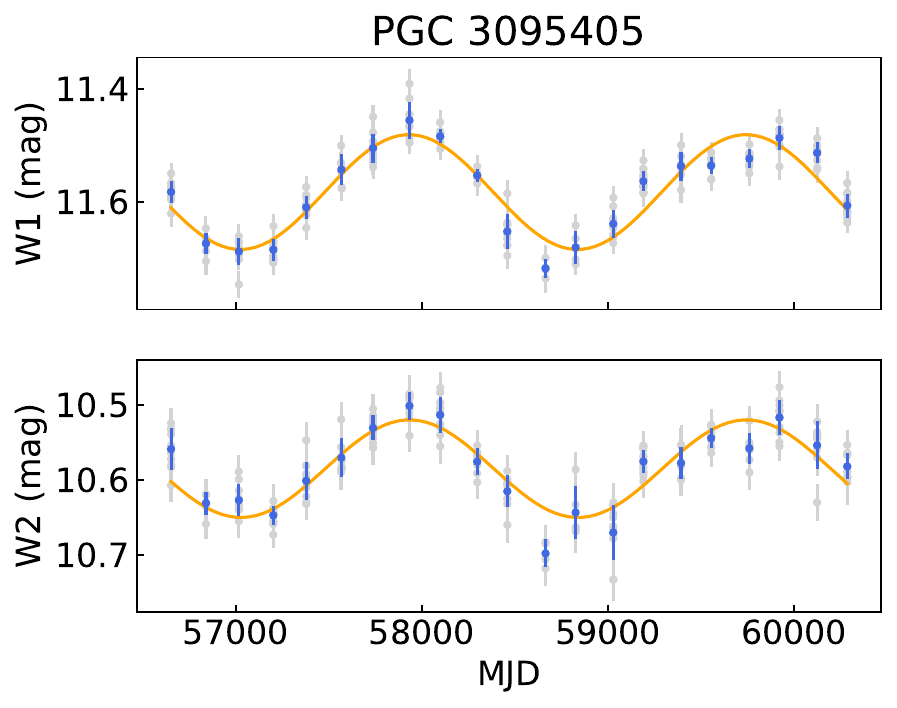}}
\centering{\includegraphics[width=0.49\textwidth]{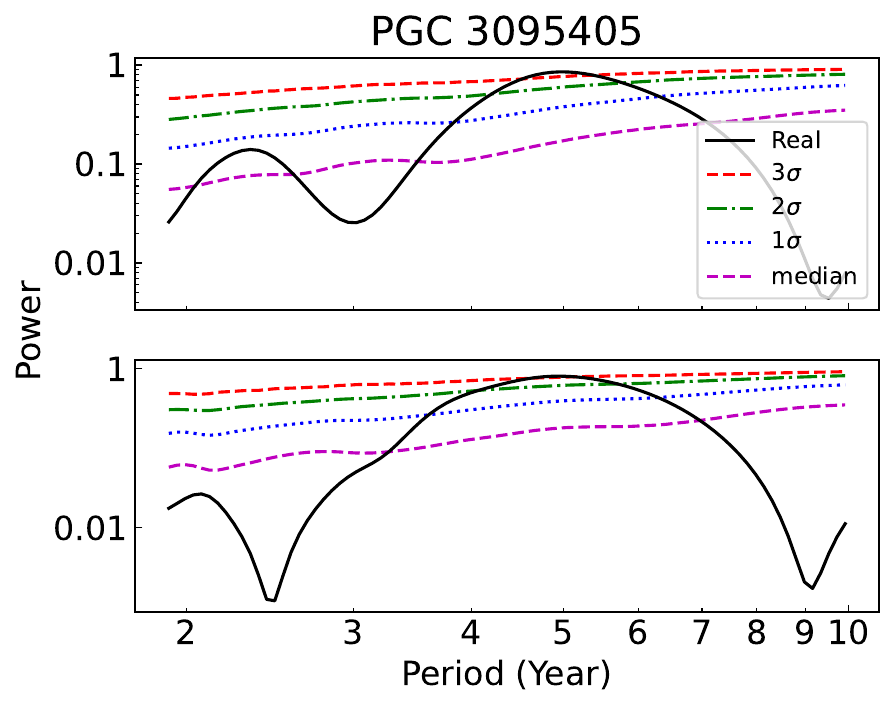}}
\end{minipage}
\begin{minipage}{1.0\textwidth}
\centering{\includegraphics[width=0.49\textwidth]{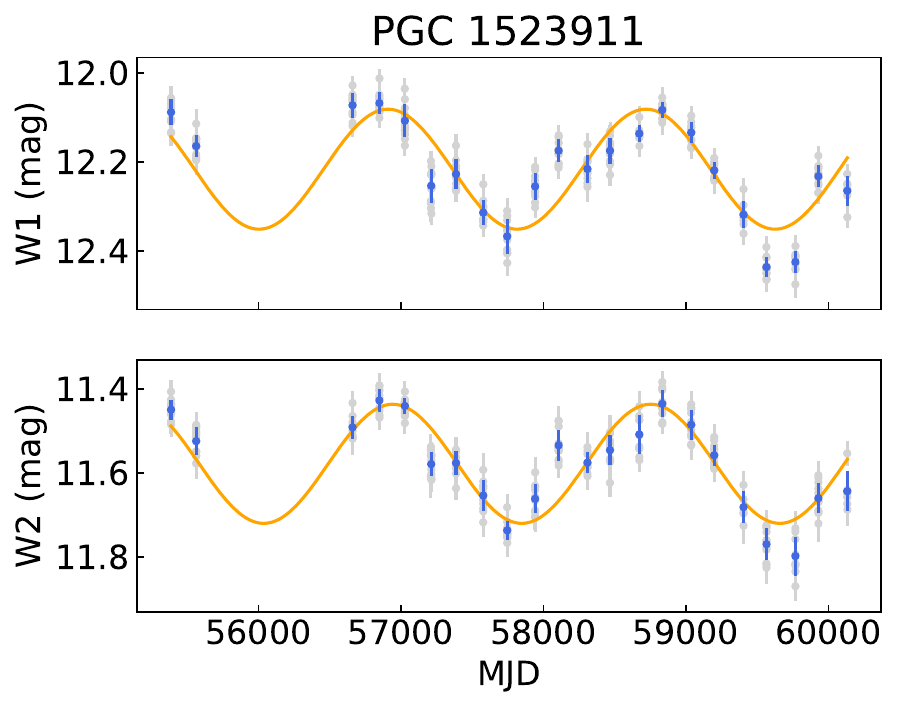}}
\centering{\includegraphics[width=0.49\textwidth]{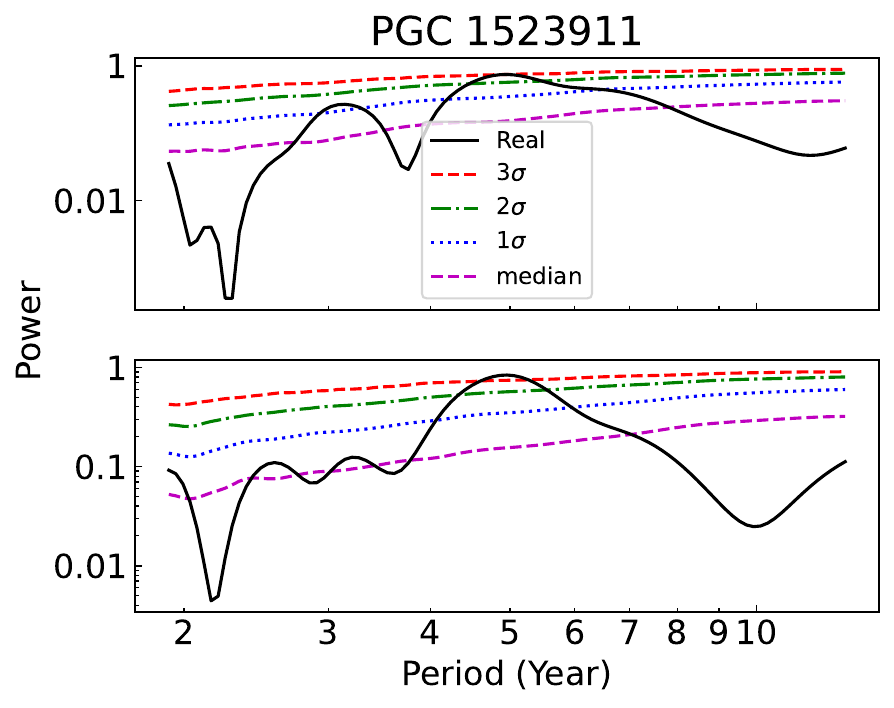}}
\end{minipage}
\begin{minipage}{1.0\textwidth}
\centering{\includegraphics[width=0.49\textwidth]{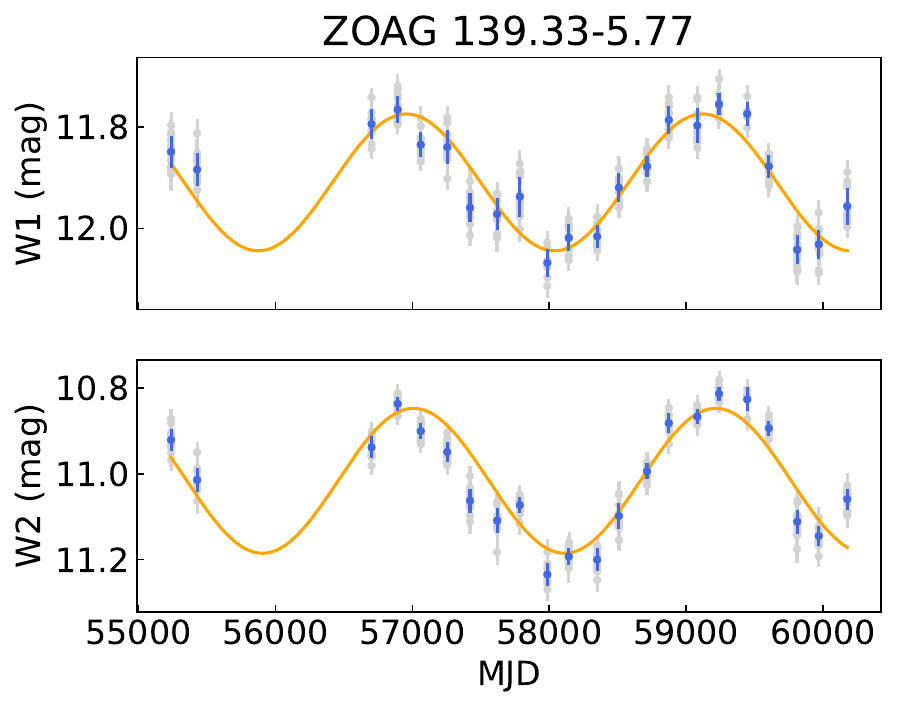}}
\centering{\includegraphics[width=0.49\textwidth]{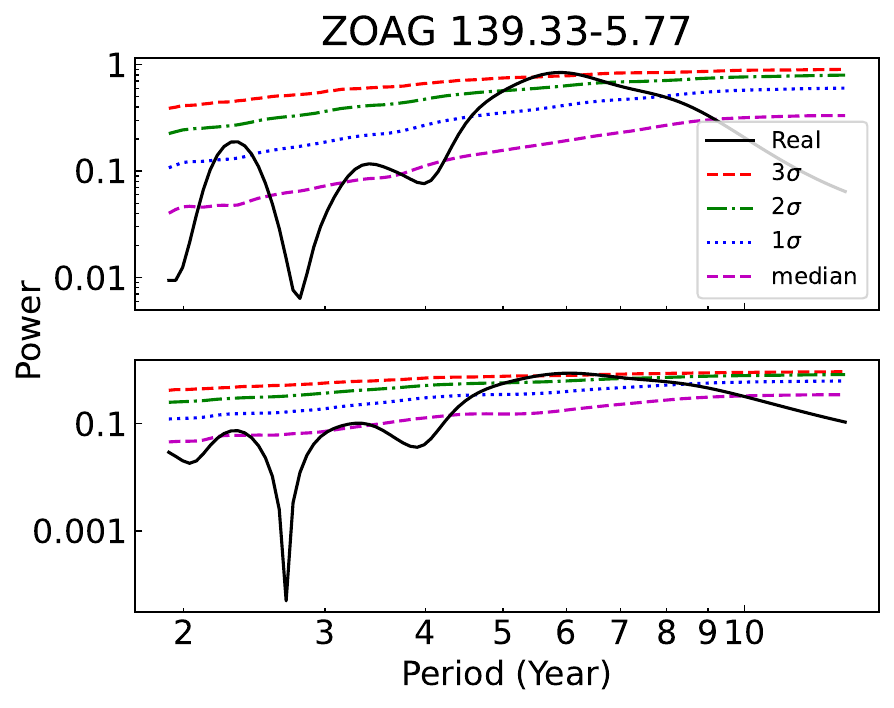}}
\end{minipage}
\caption{The light curves and periodograms of the rest 26 candidates. The legends correspond to those in Fig.~\ref{lc}.}
\label{lcr}
\end{figure*}

\begin{figure*}
\centering
\begin{minipage}{1.0\textwidth}
\centering{\includegraphics[width=0.49\textwidth]{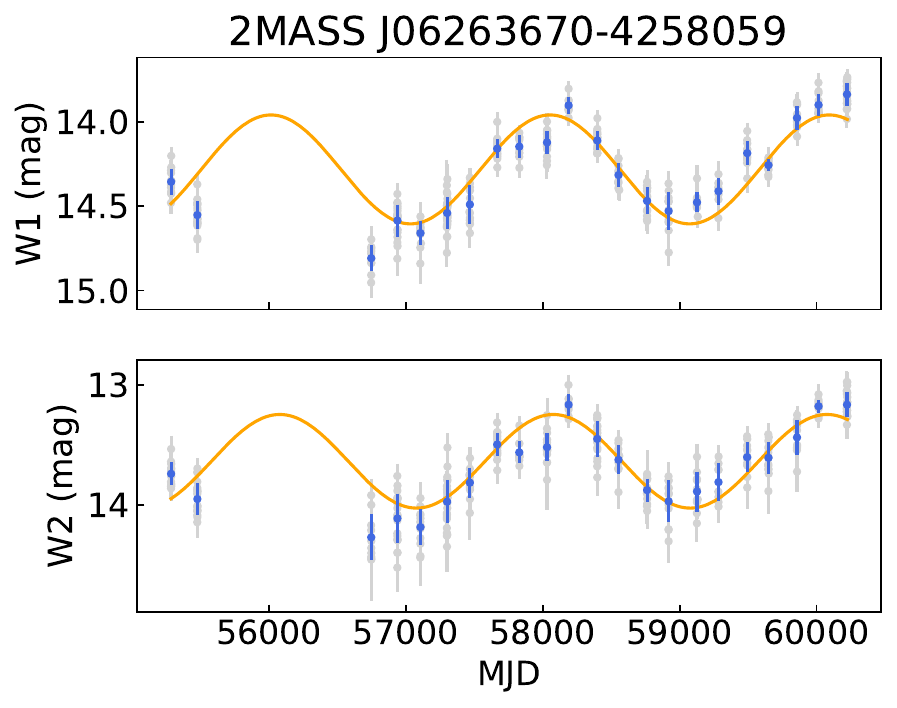}}
\centering{\includegraphics[width=0.49\textwidth]{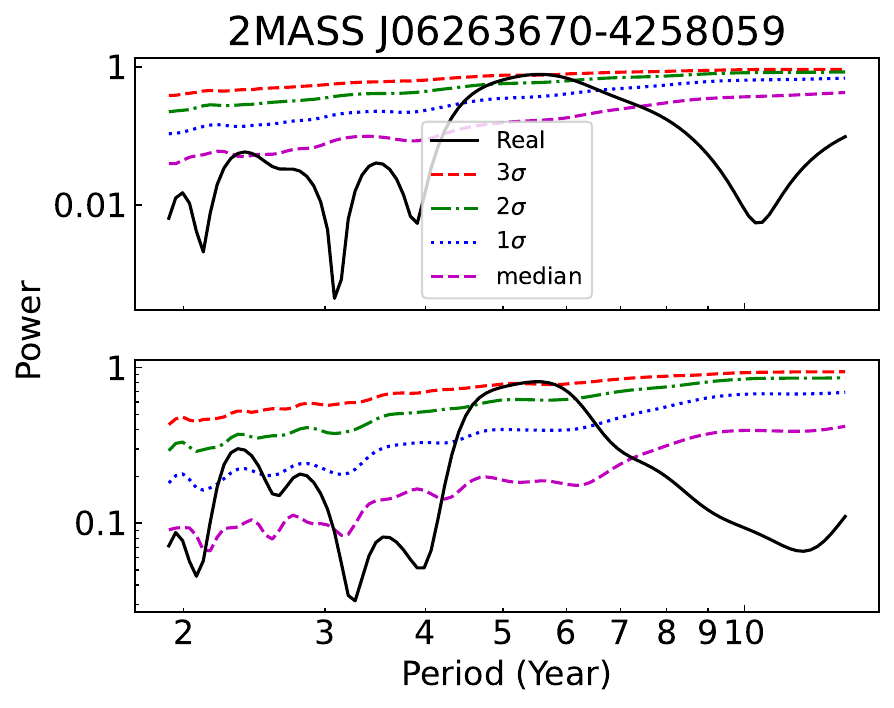}}
\end{minipage}
\begin{minipage}{1.0\textwidth}
\centering{\includegraphics[width=0.49\textwidth]{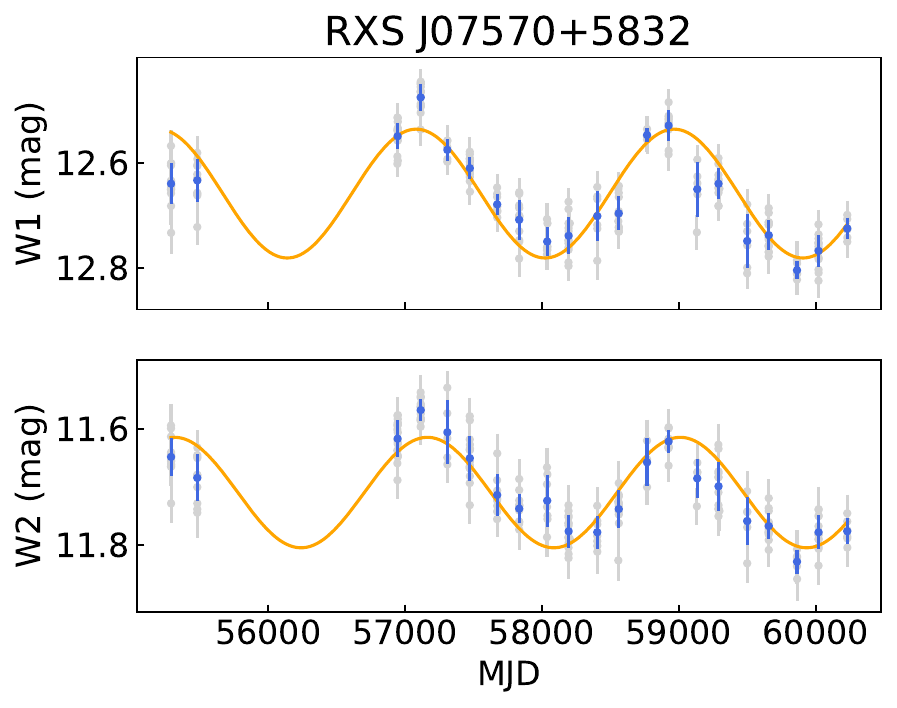}}
\centering{\includegraphics[width=0.49\textwidth]{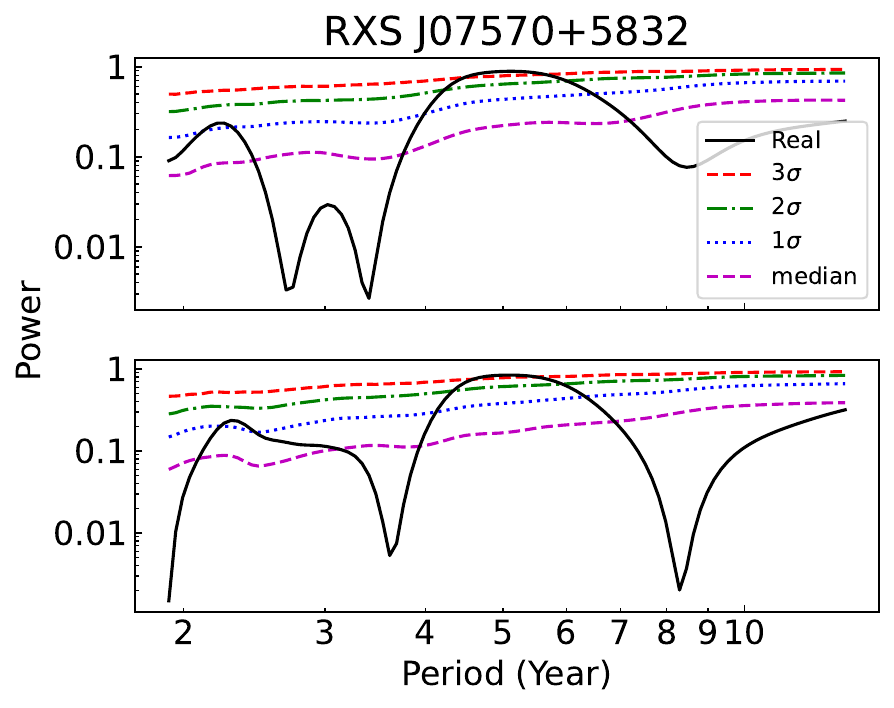}}
\end{minipage}
\begin{minipage}{1.0\textwidth}
\centering{\includegraphics[width=0.49\textwidth]{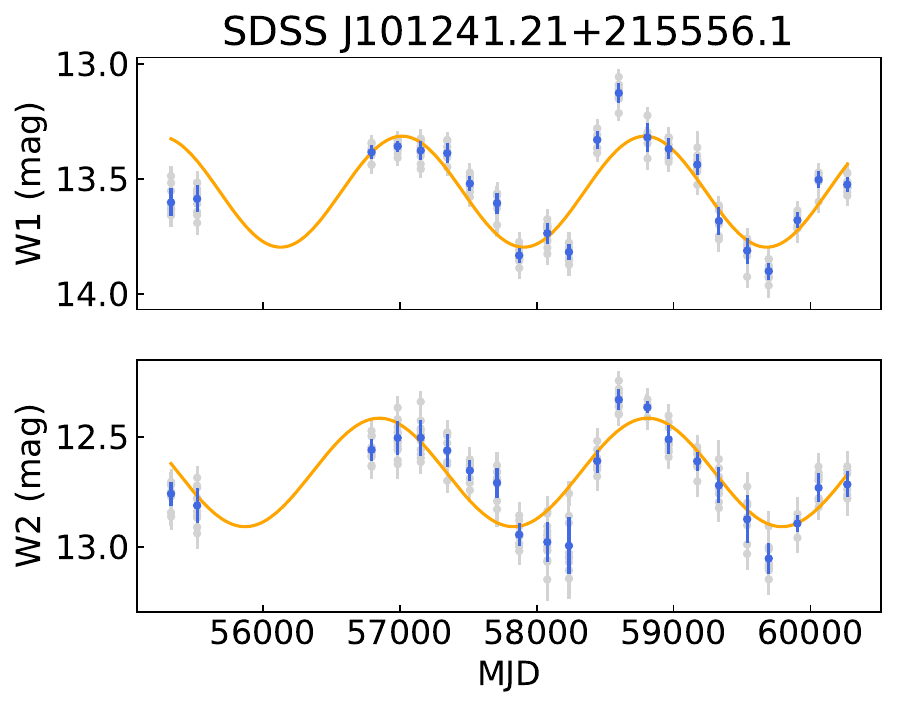}}
\centering{\includegraphics[width=0.49\textwidth]{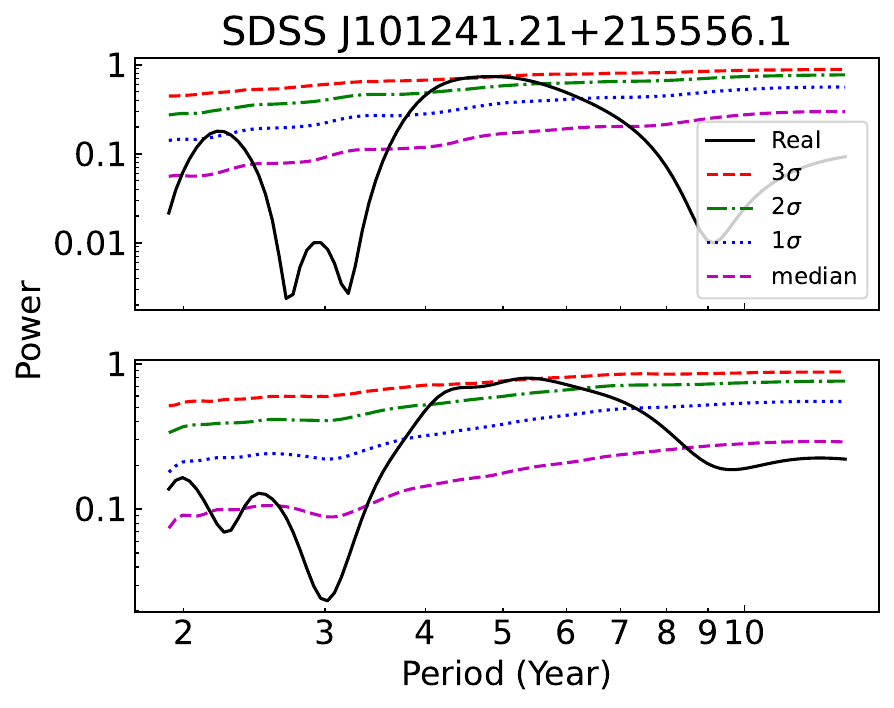}}
\end{minipage}
\addtocounter{figure}{-1}
\caption{(Continued).}
\end{figure*}

\begin{figure*}
\centering
\begin{minipage}{1.0\textwidth}
\centering{\includegraphics[width=0.49\textwidth]{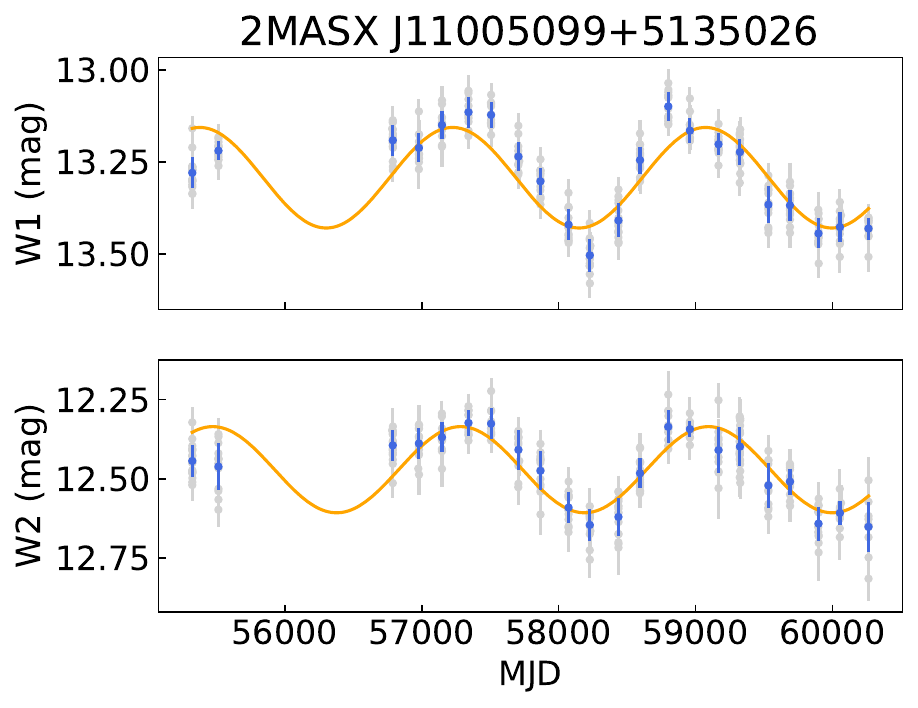}}
\centering{\includegraphics[width=0.49\textwidth]{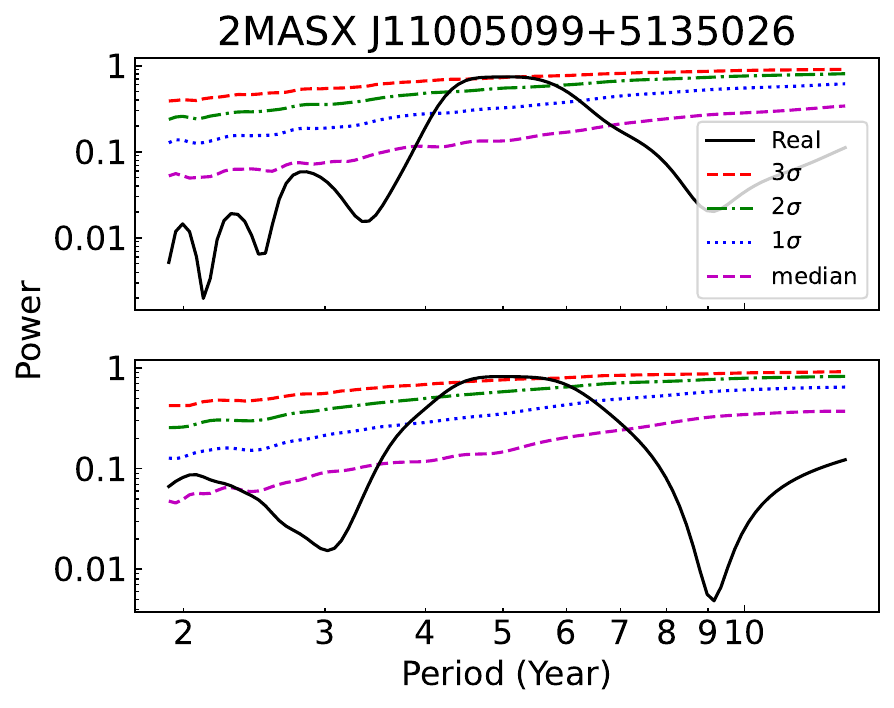}}
\end{minipage}
\begin{minipage}{1.0\textwidth}
\centering{\includegraphics[width=0.49\textwidth]{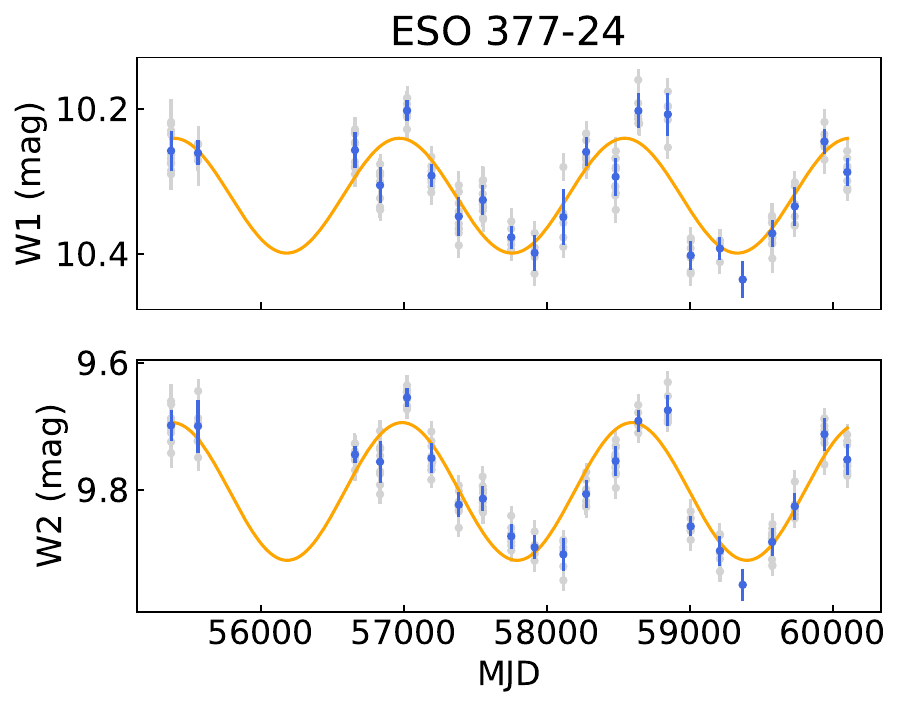}}
\centering{\includegraphics[width=0.49\textwidth]{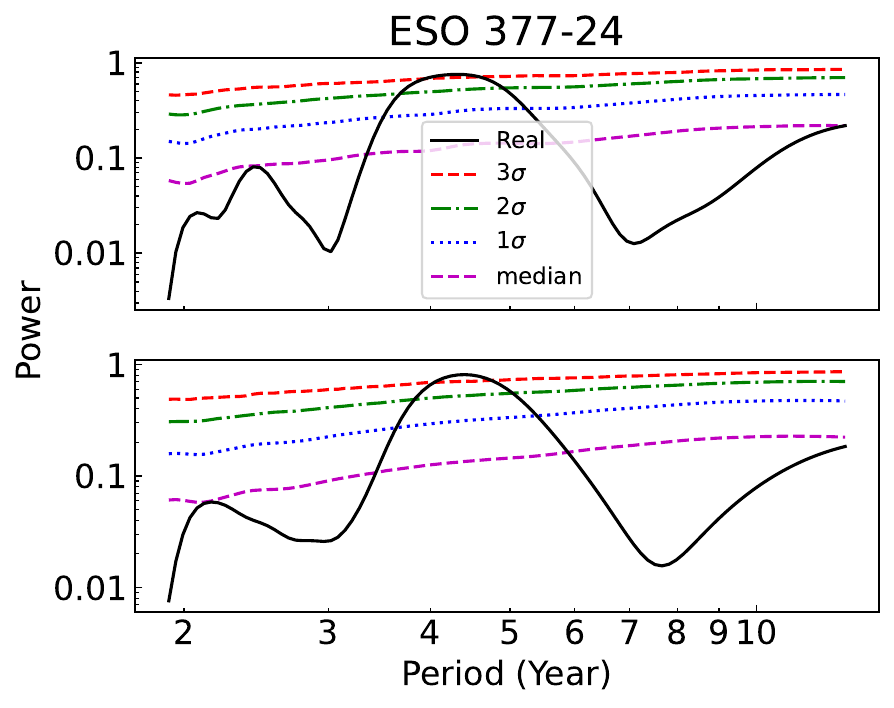}}
\end{minipage}
\begin{minipage}{1.0\textwidth}
\centering{\includegraphics[width=0.49\textwidth]{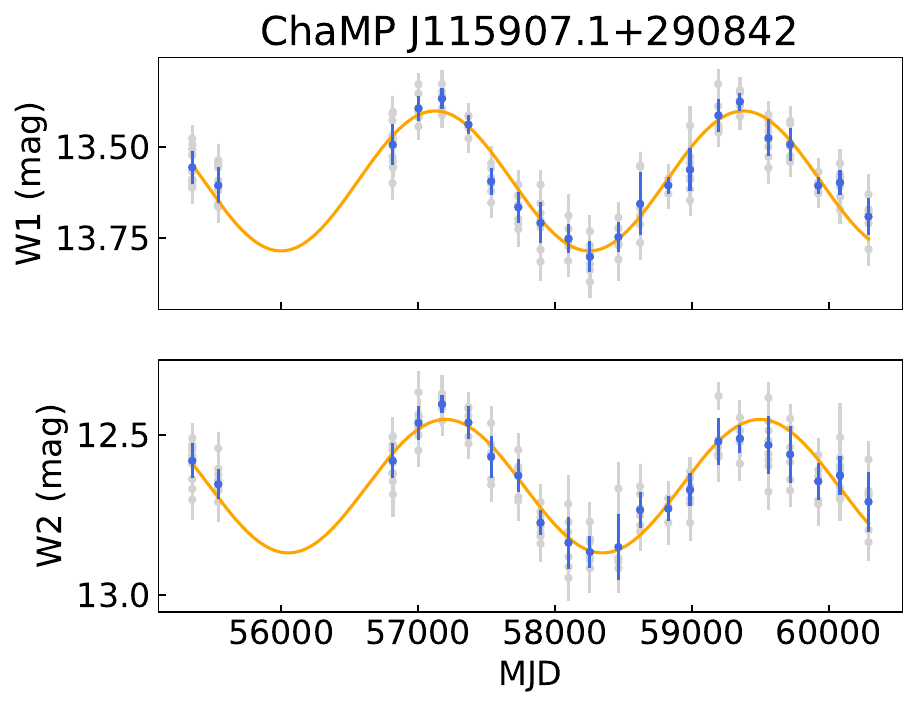}}
\centering{\includegraphics[width=0.49\textwidth]{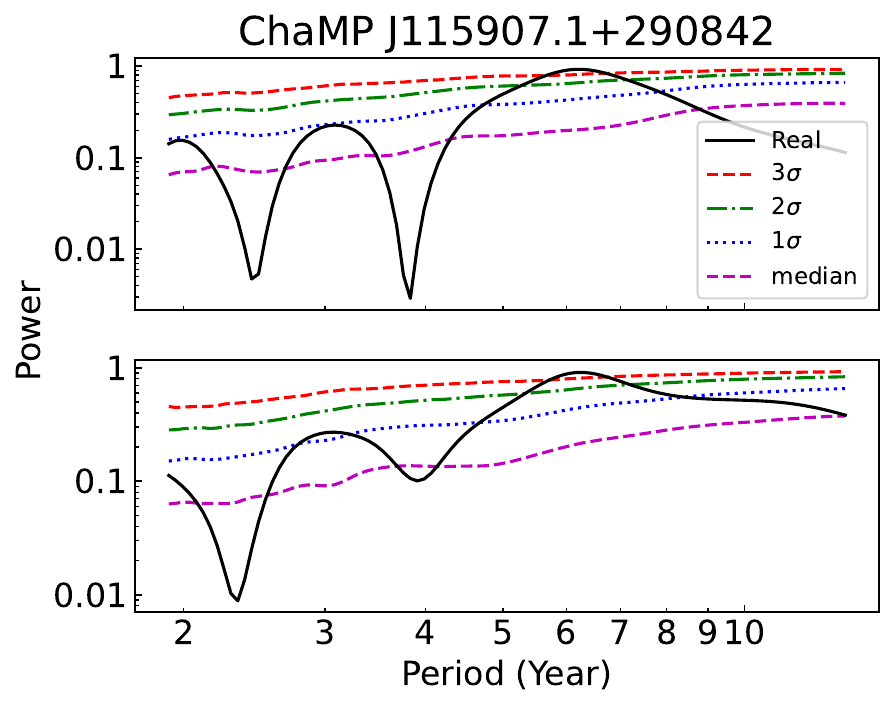}}
\end{minipage}
\addtocounter{figure}{-1}
\caption{(Continued).}
\end{figure*}

\begin{figure*}
\centering
\begin{minipage}{1.0\textwidth}
\centering{\includegraphics[width=0.49\textwidth]{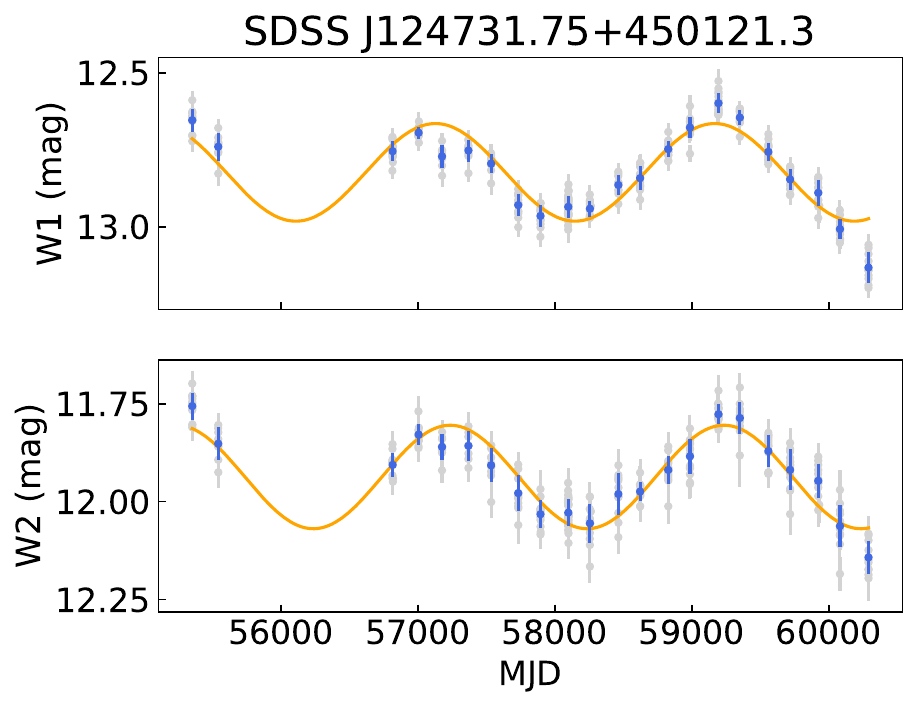}}
\centering{\includegraphics[width=0.49\textwidth]{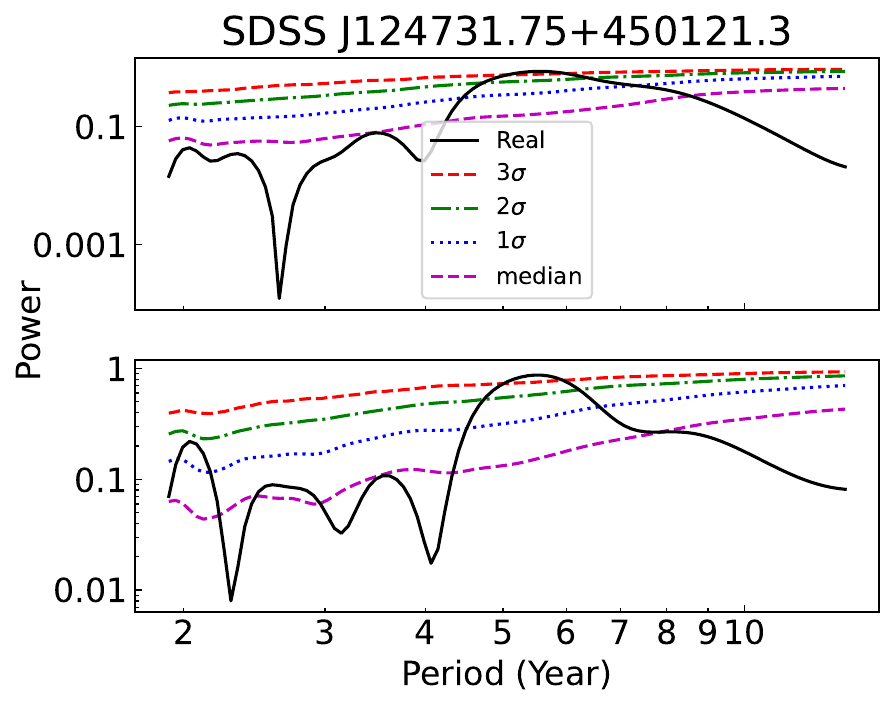}}
\end{minipage}
\begin{minipage}{1.0\textwidth}
\centering{\includegraphics[width=0.49\textwidth]{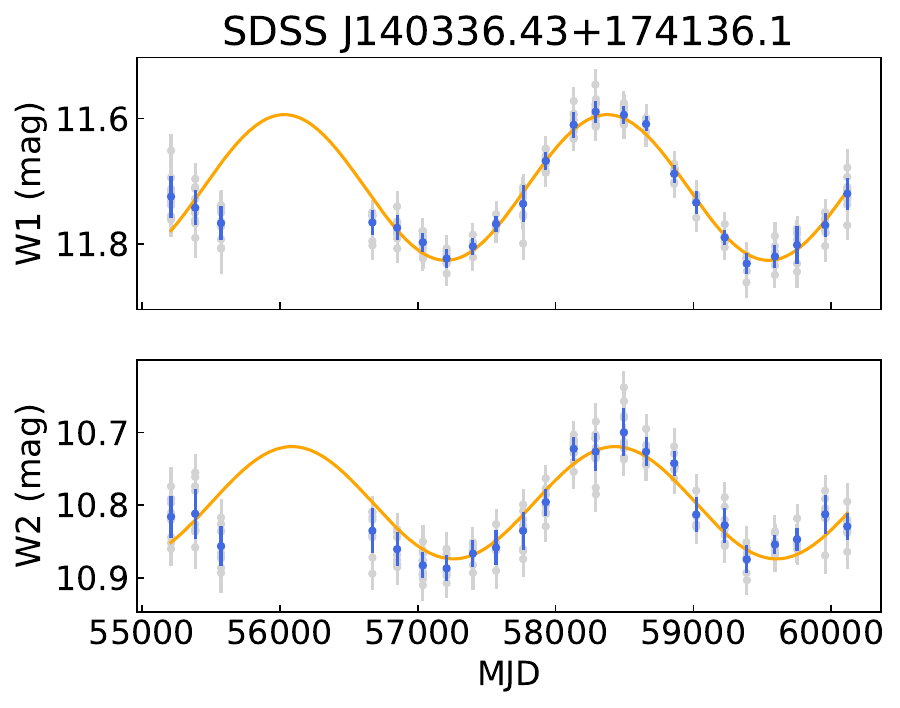}}
\centering{\includegraphics[width=0.49\textwidth]{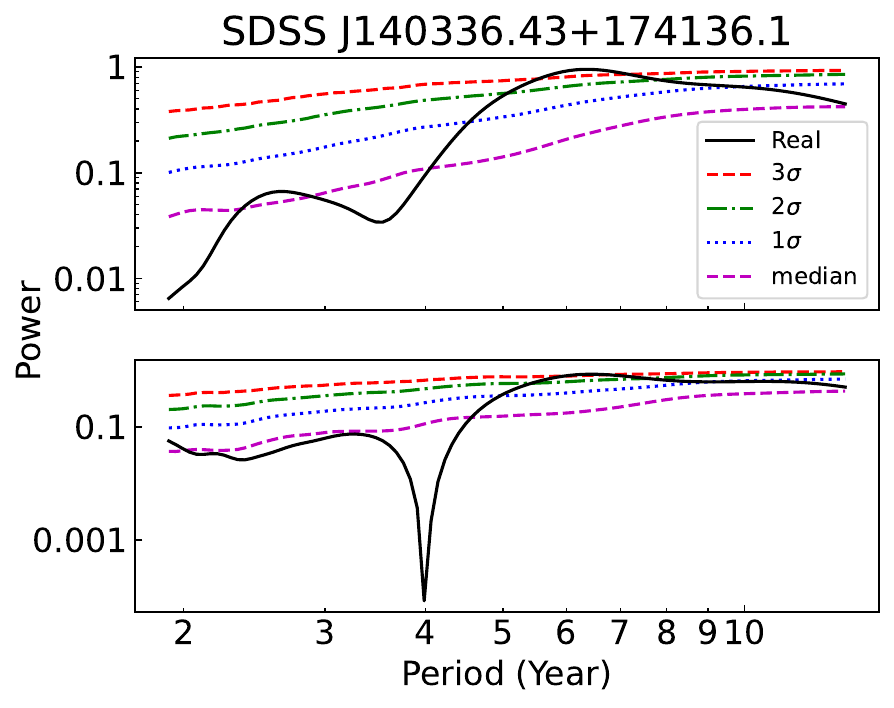}}
\end{minipage}
\begin{minipage}{1.0\textwidth}
\centering{\includegraphics[width=0.49\textwidth]{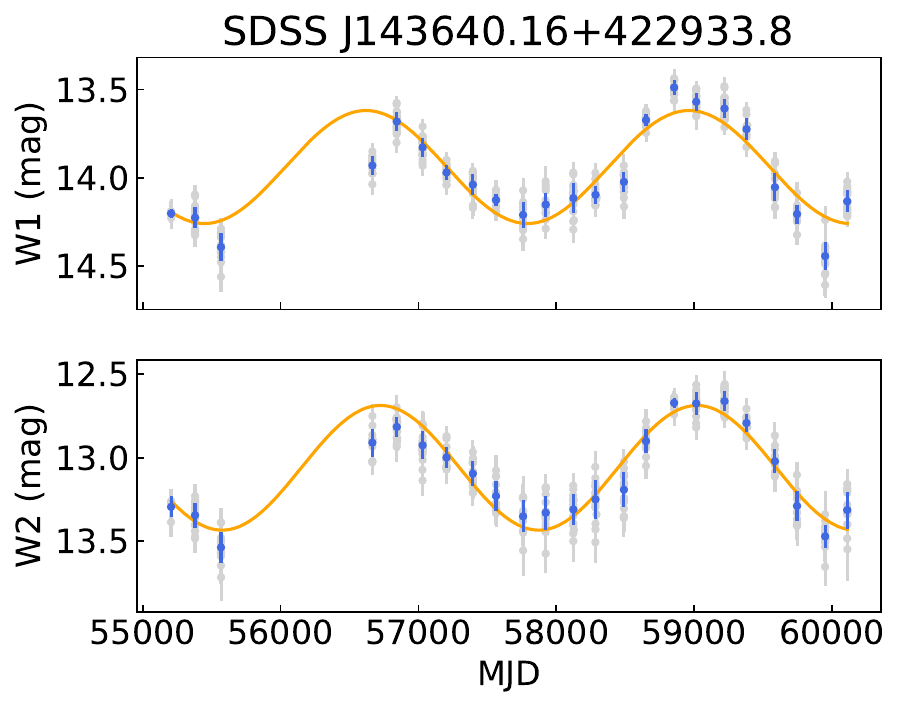}}
\centering{\includegraphics[width=0.49\textwidth]{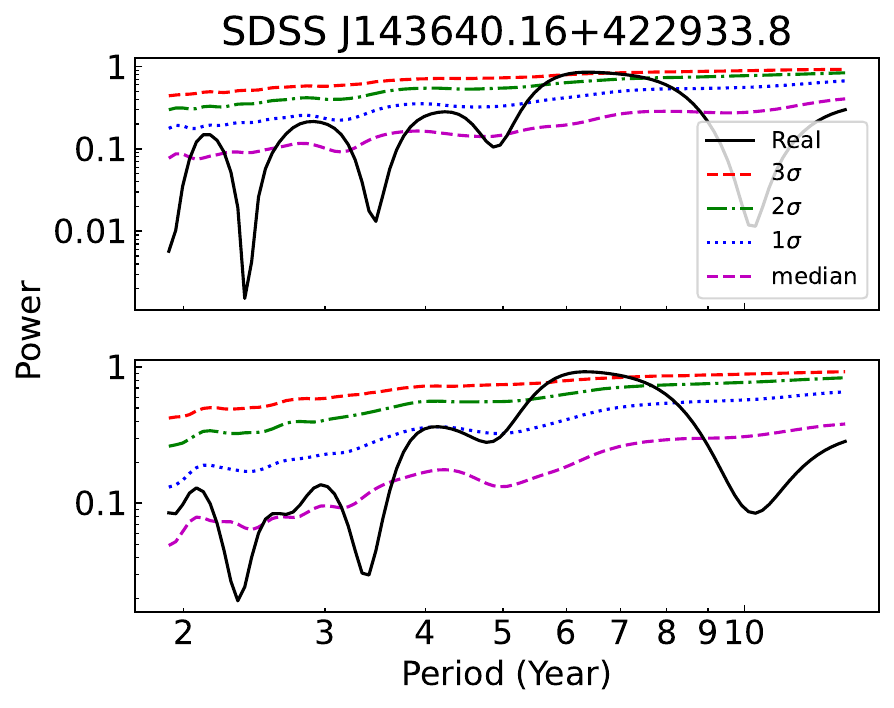}}
\end{minipage}
\addtocounter{figure}{-1}
\caption{(Continued).}
\end{figure*}

\begin{figure*}
\centering
\begin{minipage}{1.0\textwidth}
\centering{\includegraphics[width=0.49\textwidth]{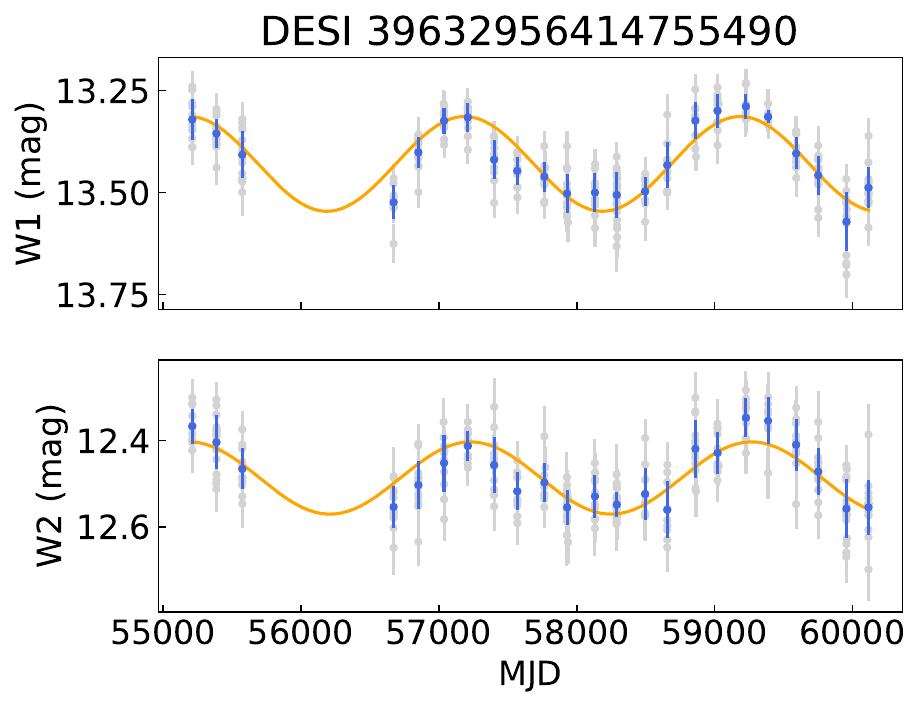}}
\centering{\includegraphics[width=0.49\textwidth]{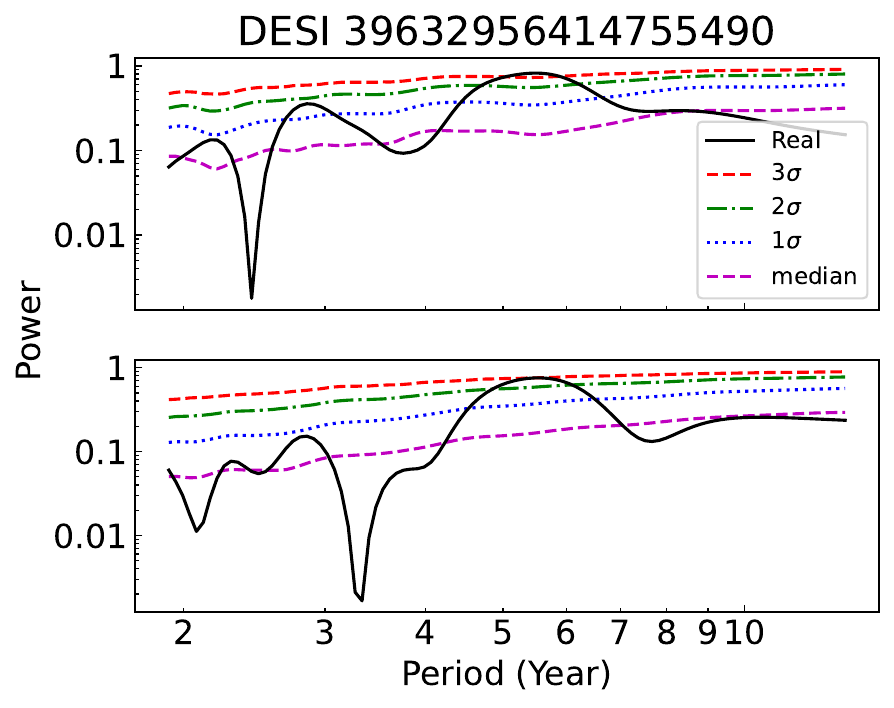}}
\end{minipage}
\begin{minipage}{1.0\textwidth}
\centering{\includegraphics[width=0.49\textwidth]{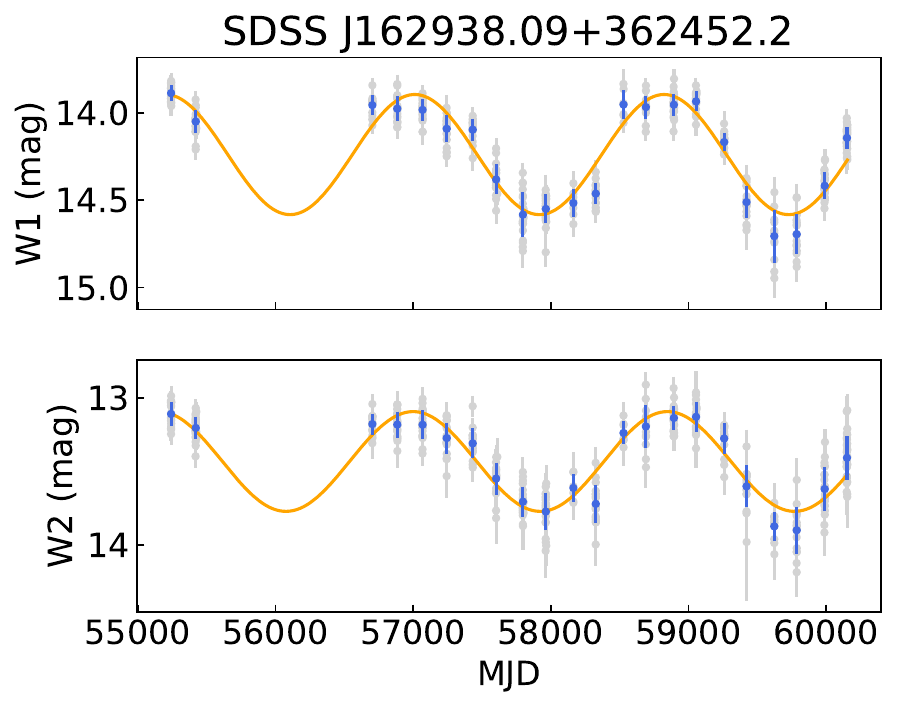}}
\centering{\includegraphics[width=0.49\textwidth]{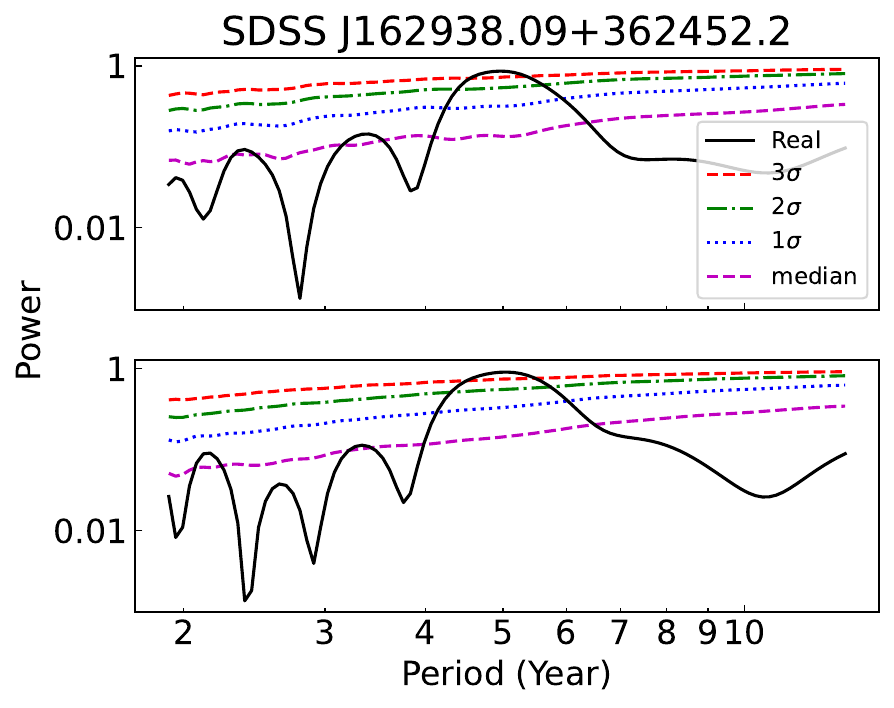}}
\end{minipage}
\begin{minipage}{1.0\textwidth}
\centering{\includegraphics[width=0.49\textwidth]{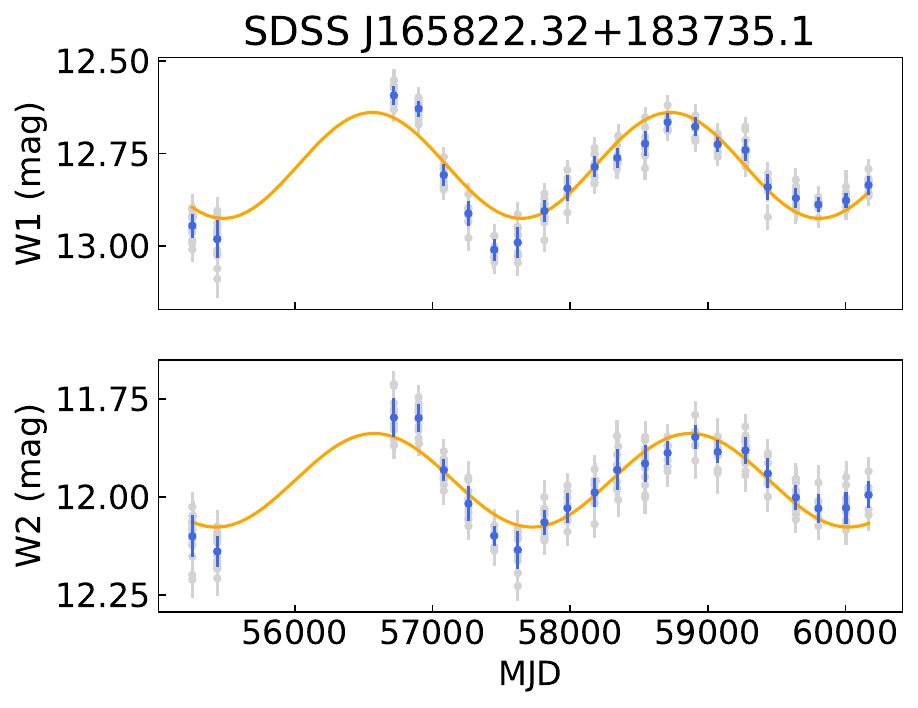}}
\centering{\includegraphics[width=0.49\textwidth]{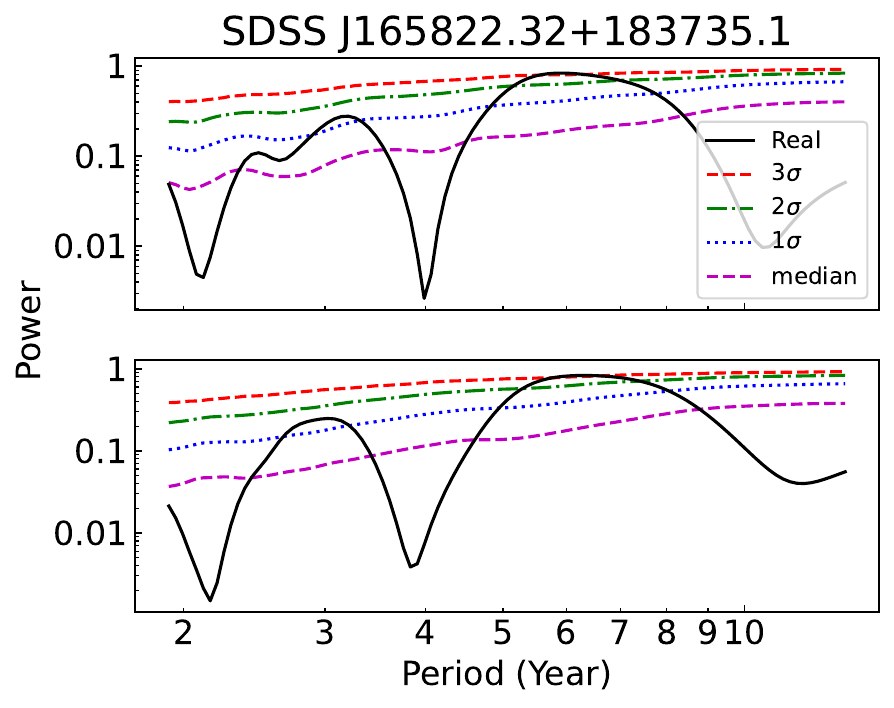}}
\end{minipage}
\addtocounter{figure}{-1}
\caption{(Continued).}
\end{figure*}

\begin{figure*}
\centering
\begin{minipage}{1.0\textwidth}
\centering{\includegraphics[width=0.49\textwidth]{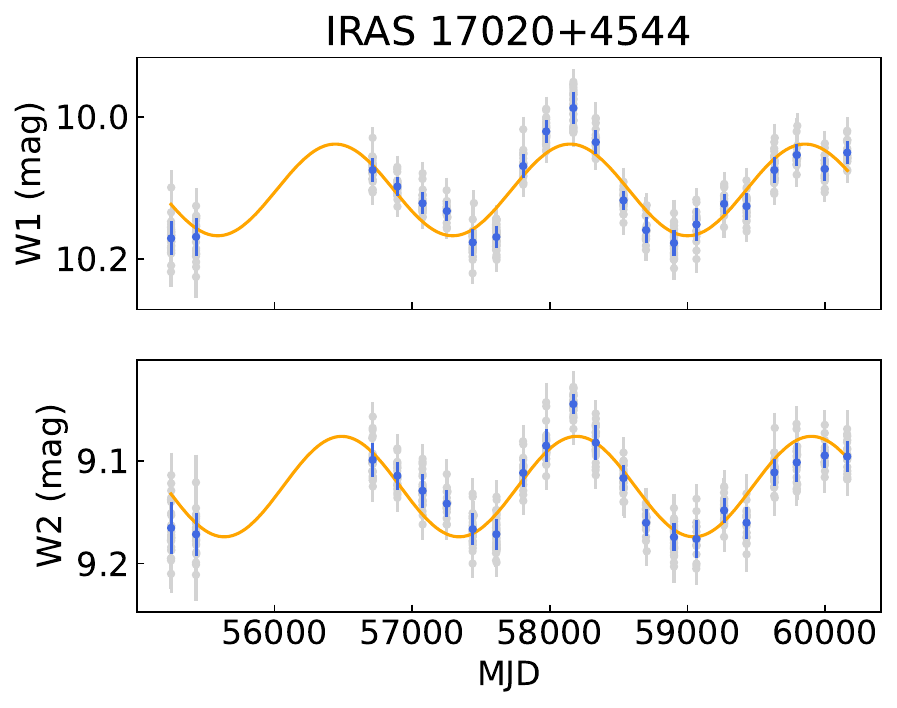}}
\centering{\includegraphics[width=0.49\textwidth]{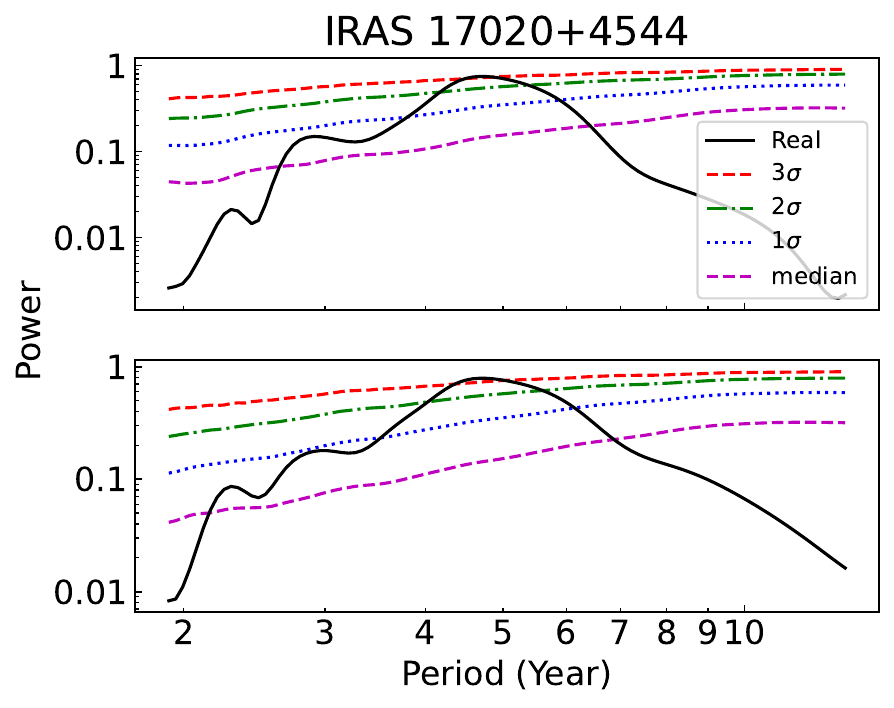}}
\end{minipage}
\begin{minipage}{1.0\textwidth}
\centering{\includegraphics[width=0.49\textwidth]{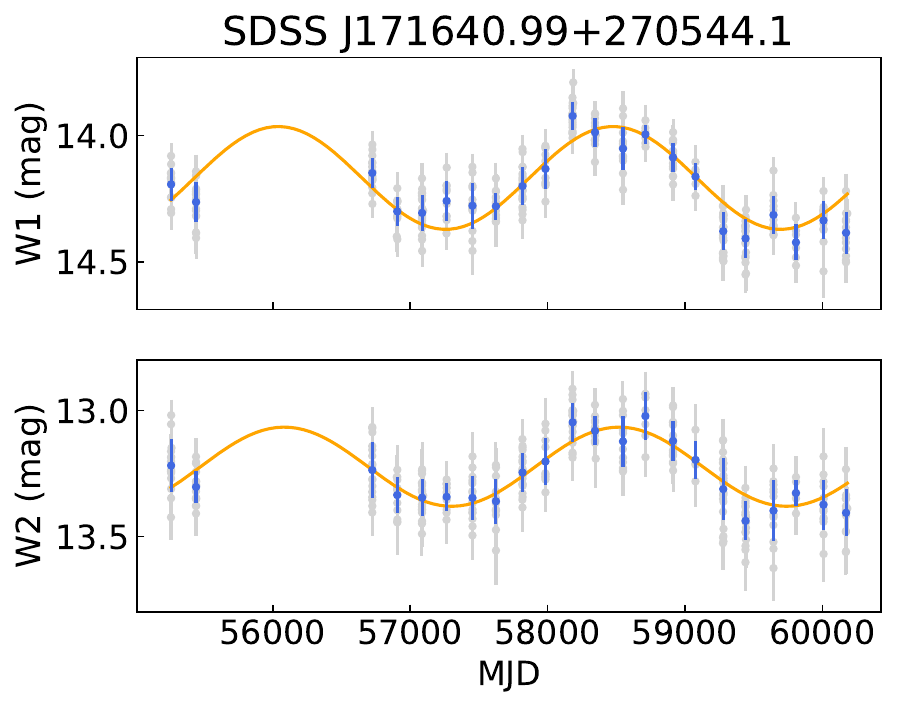}}
\centering{\includegraphics[width=0.49\textwidth]{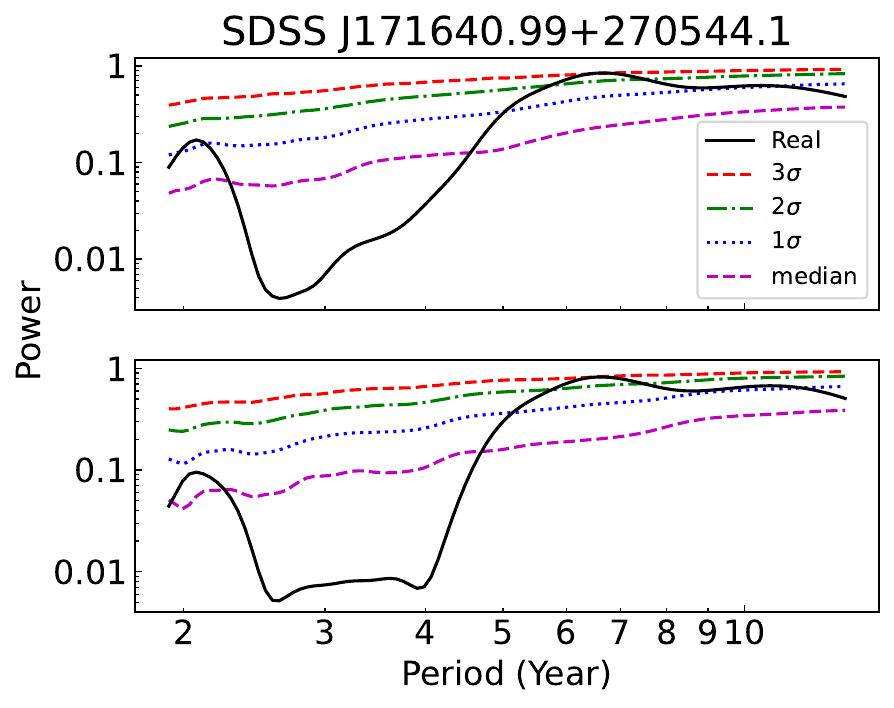}}
\end{minipage}
\begin{minipage}{1.0\textwidth}
\centering{\includegraphics[width=0.49\textwidth]{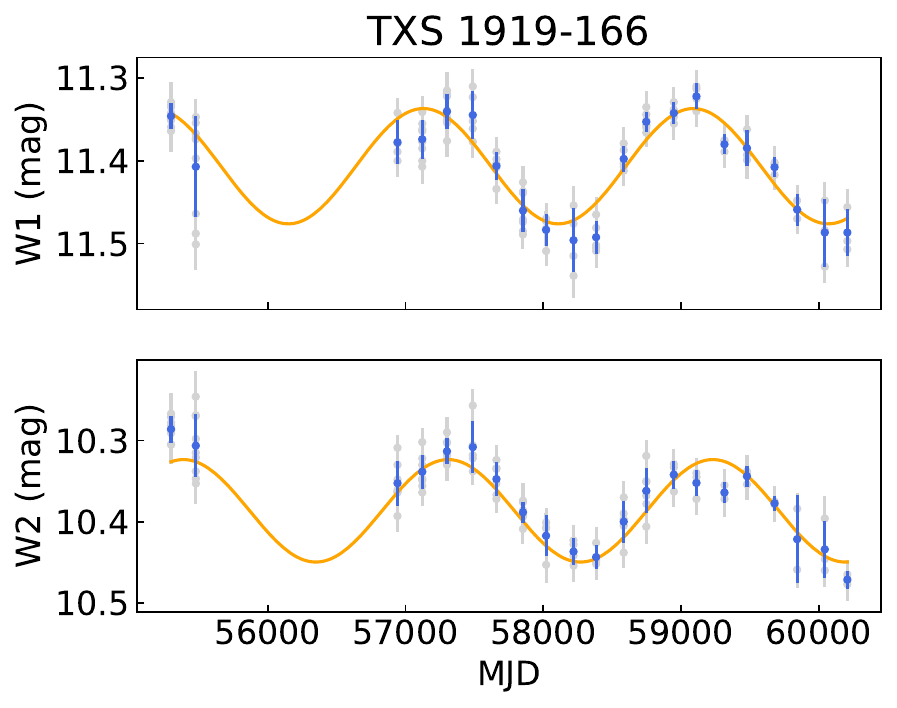}}
\centering{\includegraphics[width=0.49\textwidth]{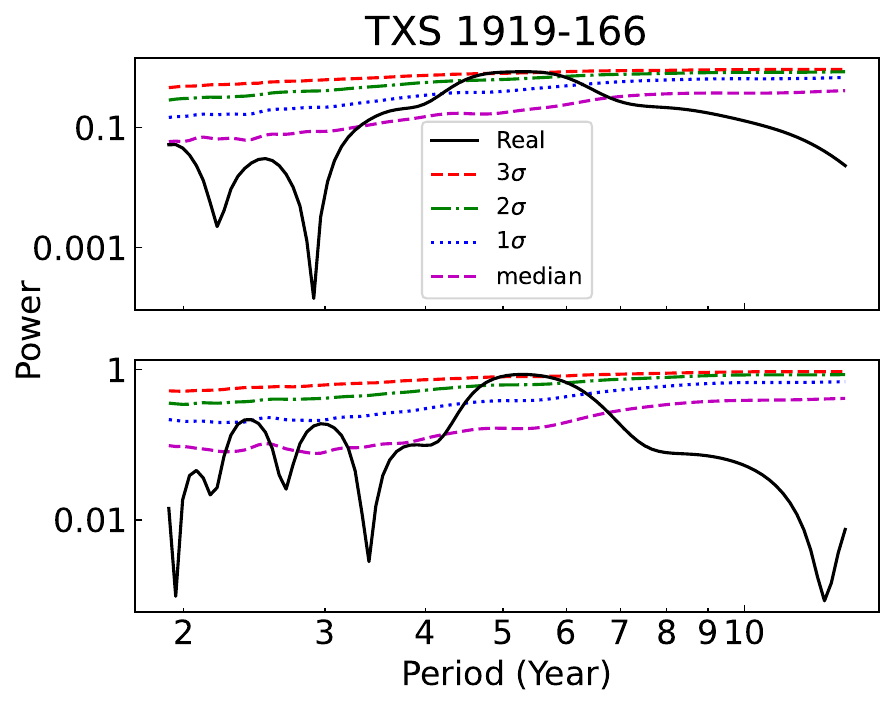}}
\end{minipage}
\addtocounter{figure}{-1}
\caption{(Continued).}
\end{figure*}

\begin{figure*}
\centering
\begin{minipage}{1.0\textwidth}
\centering{\includegraphics[width=0.49\textwidth]{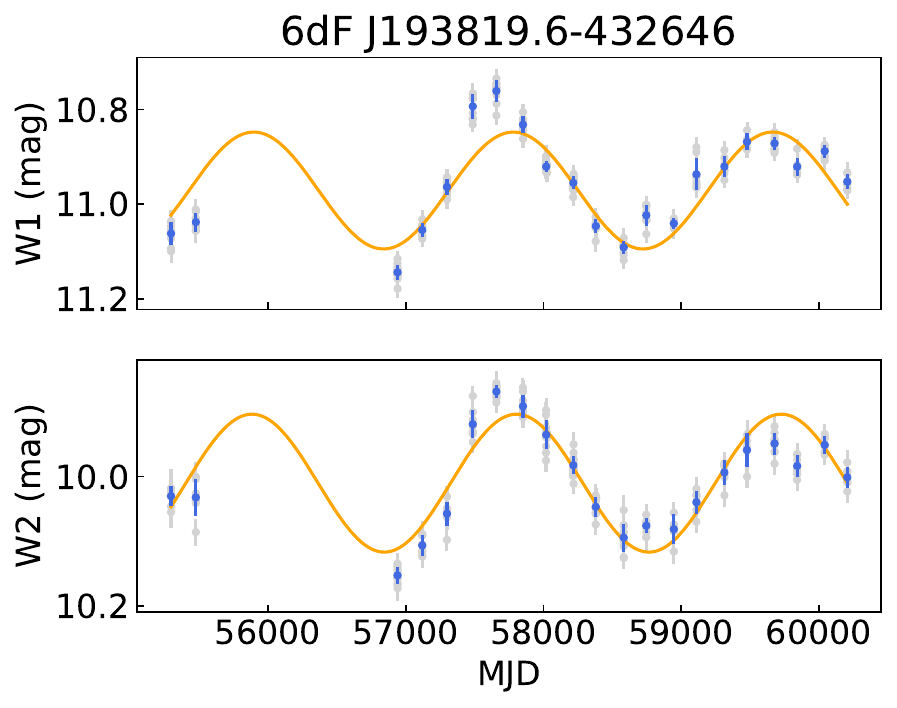}}
\centering{\includegraphics[width=0.49\textwidth]{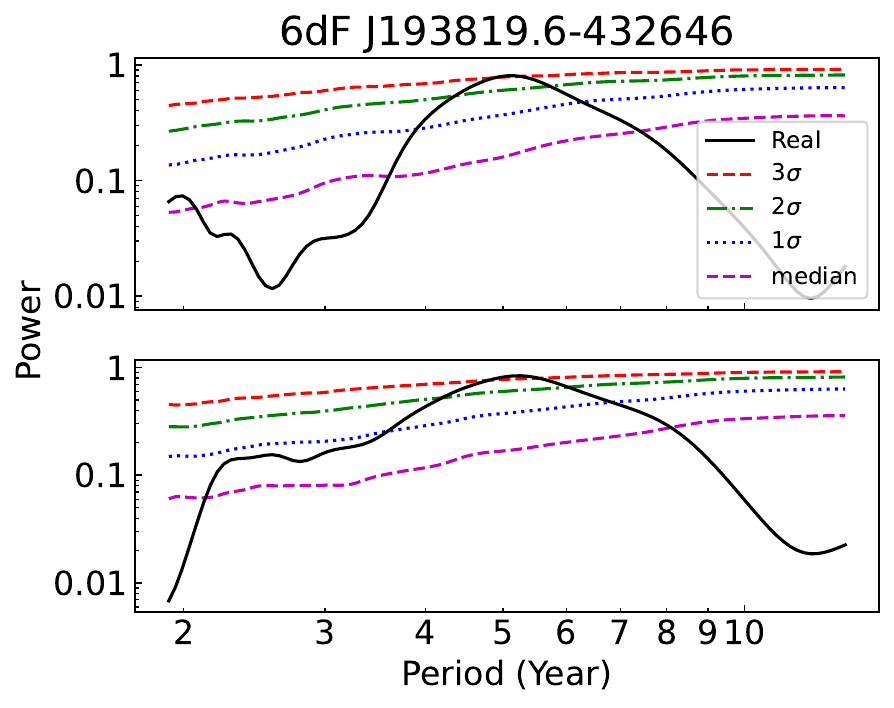}}
\end{minipage}
\begin{minipage}{1.0\textwidth}
\centering{\includegraphics[width=0.49\textwidth]{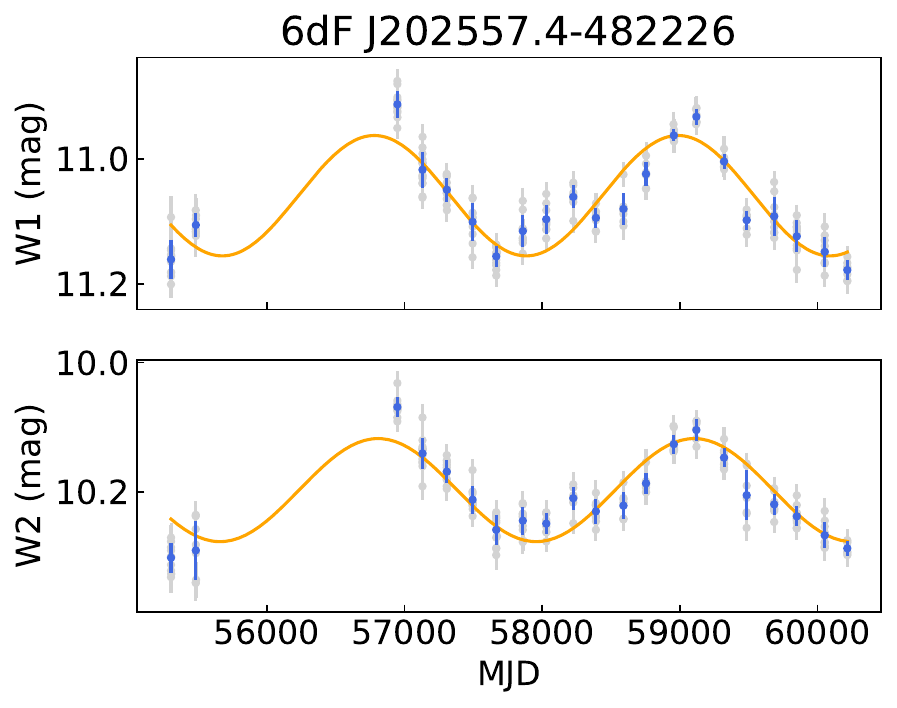}}
\centering{\includegraphics[width=0.49\textwidth]{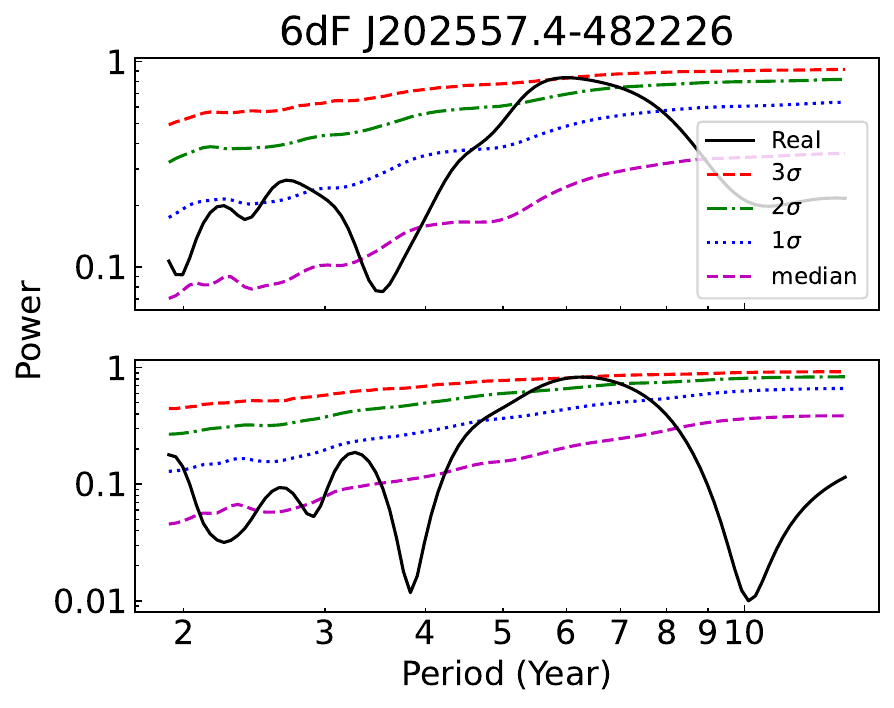}}
\end{minipage}
\begin{minipage}{1.0\textwidth}
\centering{\includegraphics[width=0.49\textwidth]{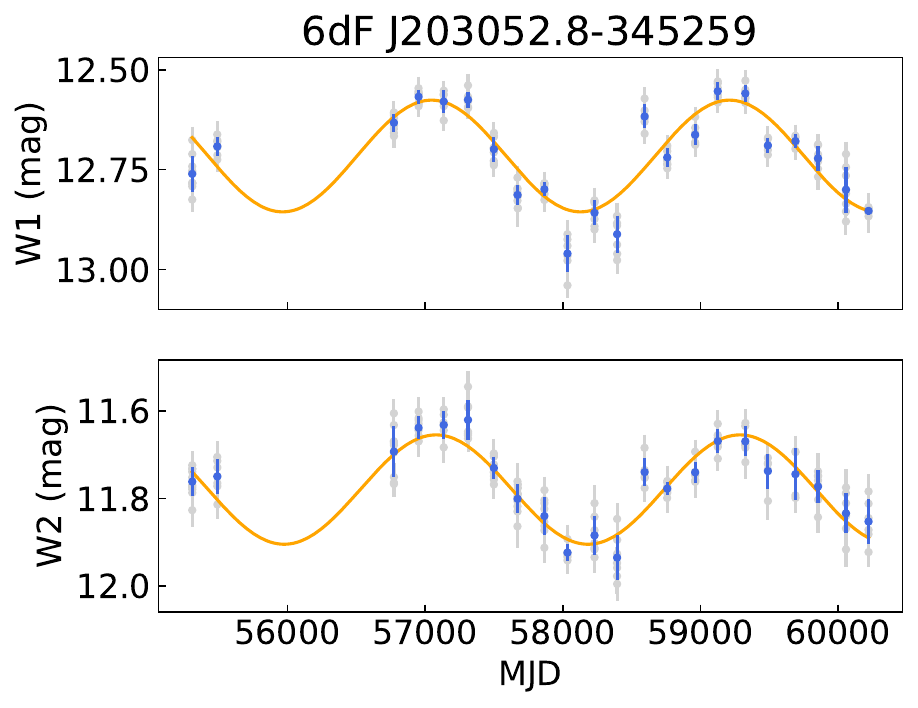}}
\centering{\includegraphics[width=0.49\textwidth]{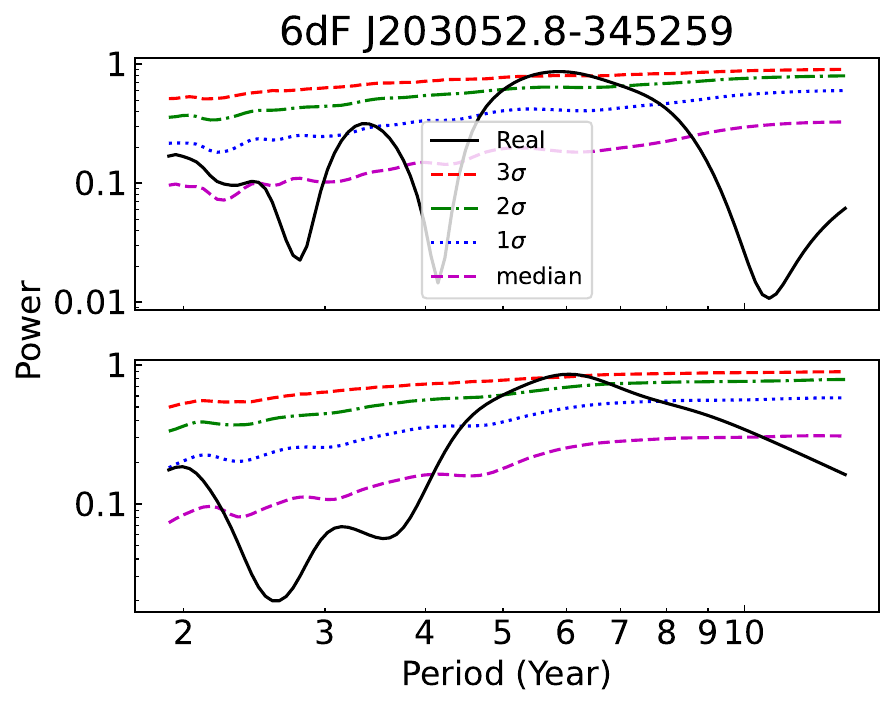}}
\end{minipage}
\addtocounter{figure}{-1}
\caption{(Continued).}
\end{figure*}

\begin{figure*}
\centering
\begin{minipage}{1.0\textwidth}
\centering{\includegraphics[width=0.49\textwidth]{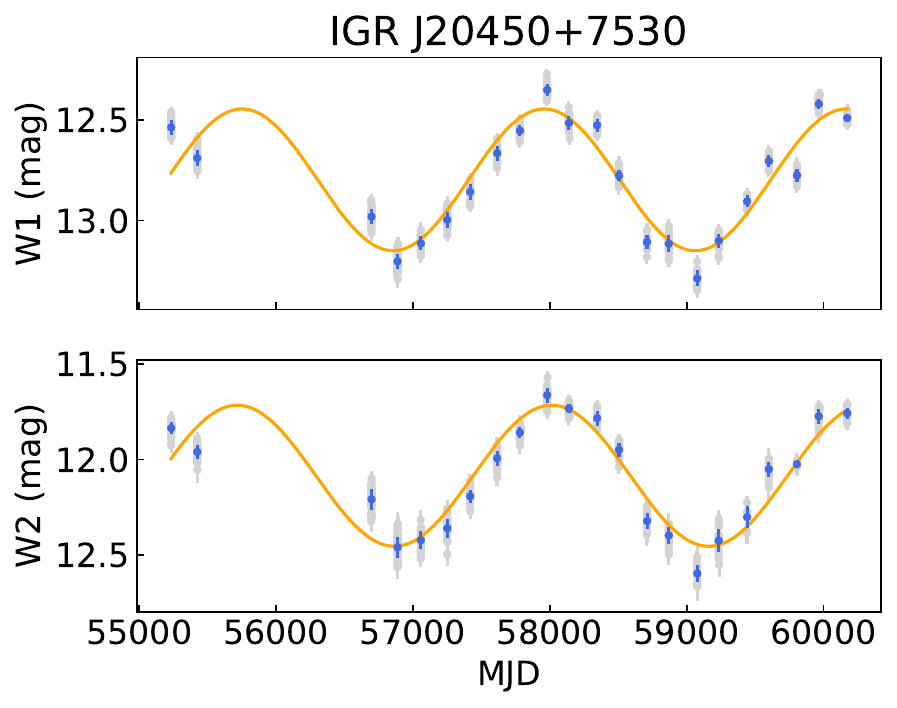}}
\centering{\includegraphics[width=0.49\textwidth]{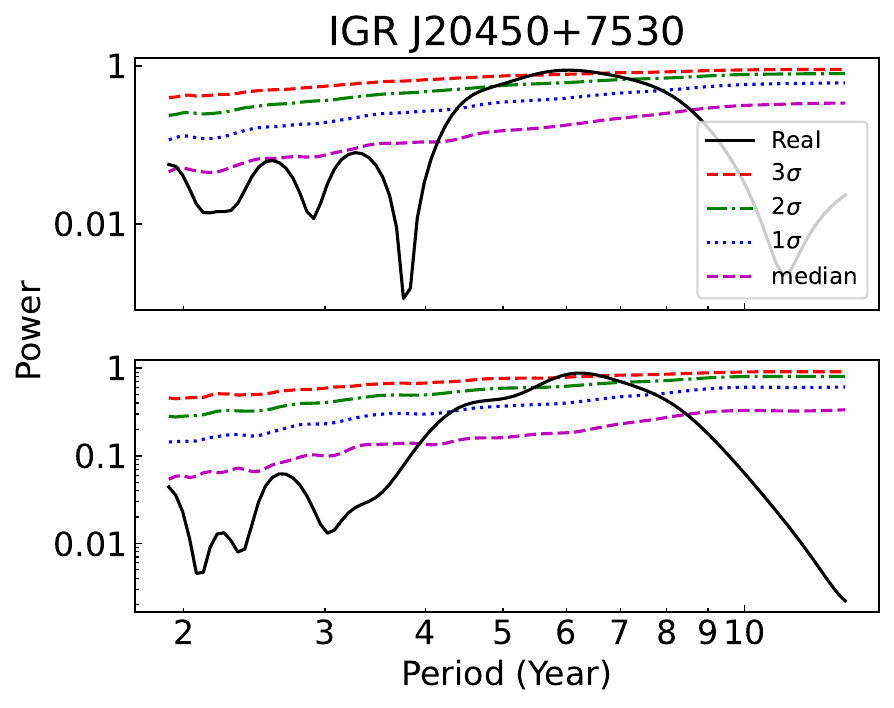}}
\end{minipage}
\begin{minipage}{1.0\textwidth}
\centering{\includegraphics[width=0.49\textwidth]{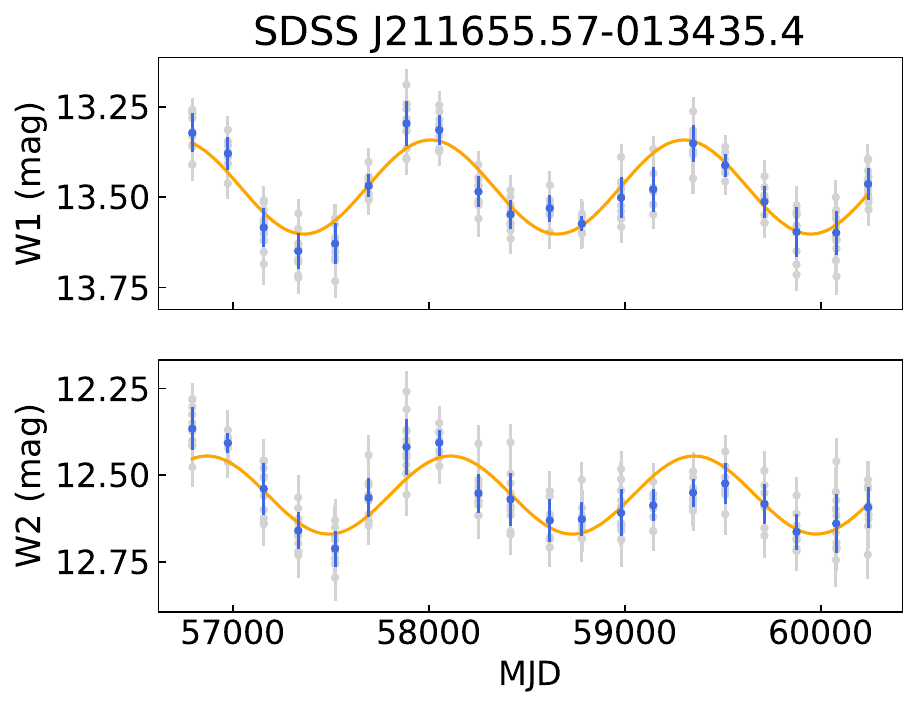}}
\centering{\includegraphics[width=0.49\textwidth]{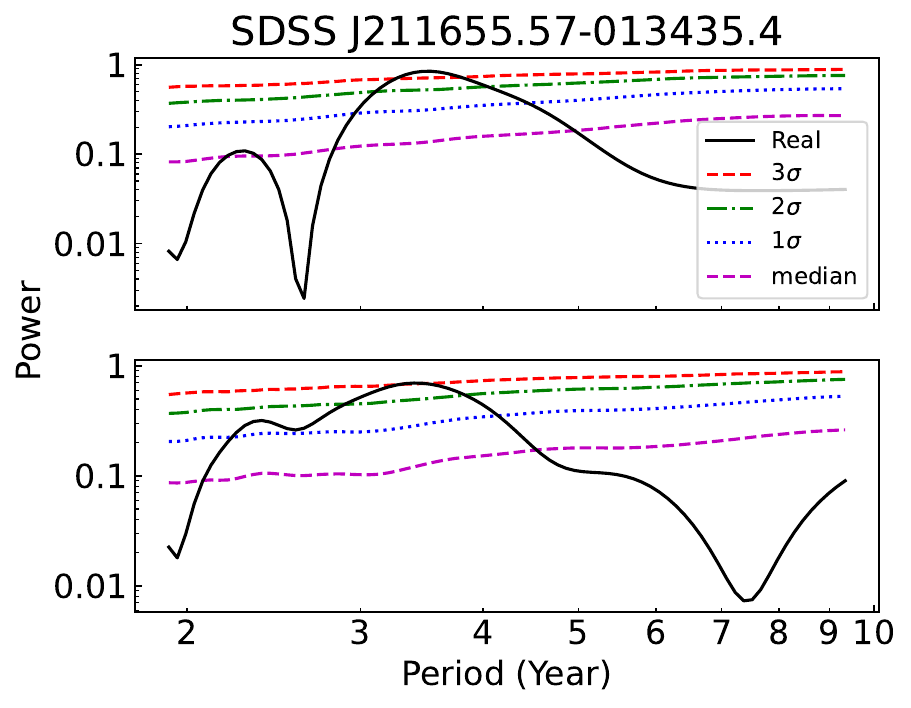}}
\end{minipage}
\begin{minipage}{1.0\textwidth}
\centering{\includegraphics[width=0.49\textwidth]{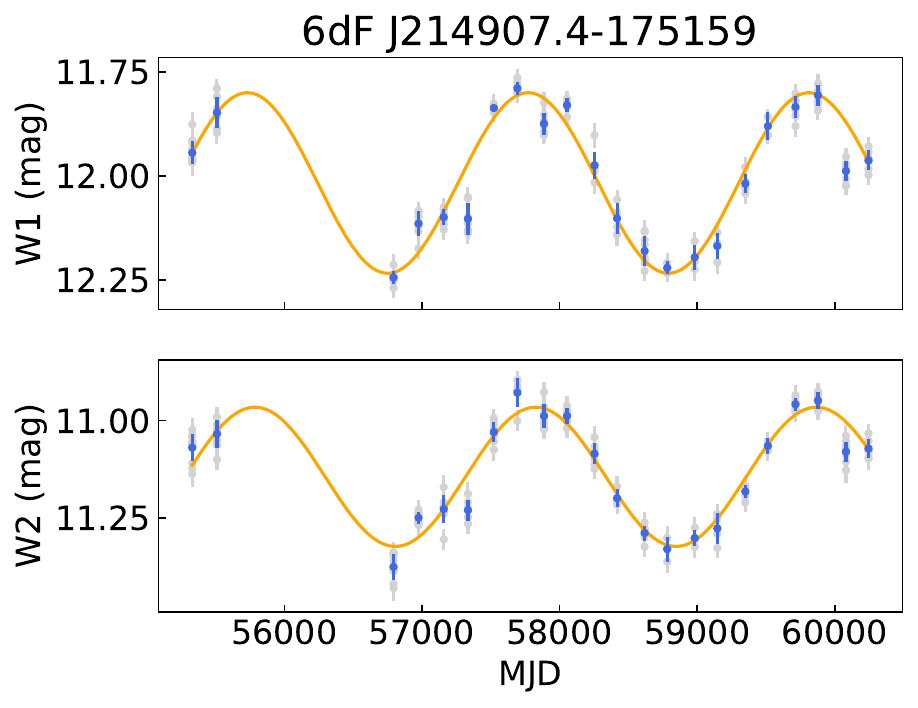}}
\centering{\includegraphics[width=0.49\textwidth]{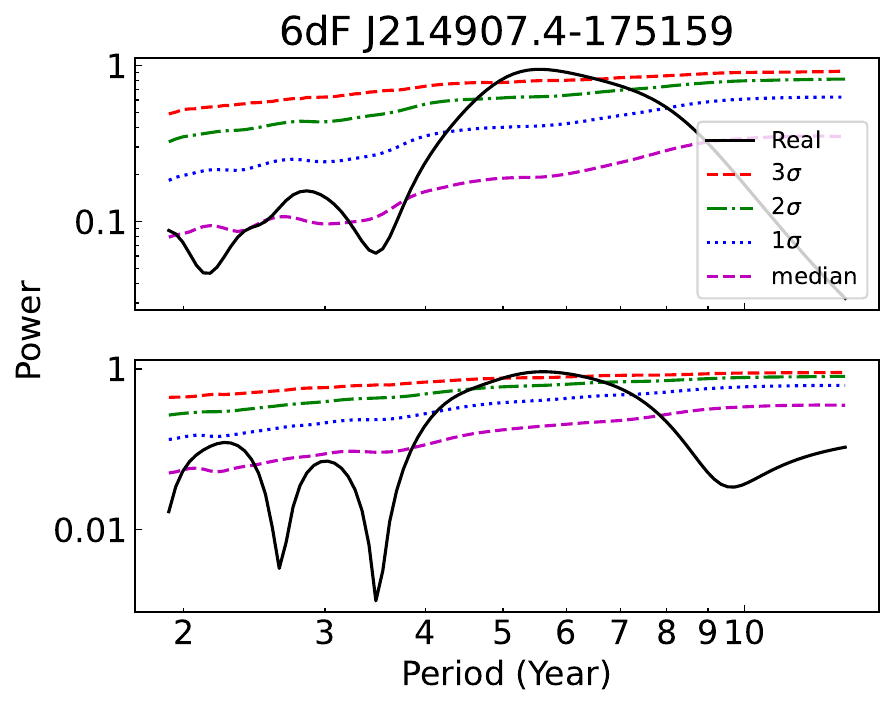}}
\end{minipage}
\addtocounter{figure}{-1}
\caption{(Continued).}
\end{figure*}

\begin{figure*}
\centering
\begin{minipage}{1.0\textwidth}
\centering{\includegraphics[width=0.49\textwidth]{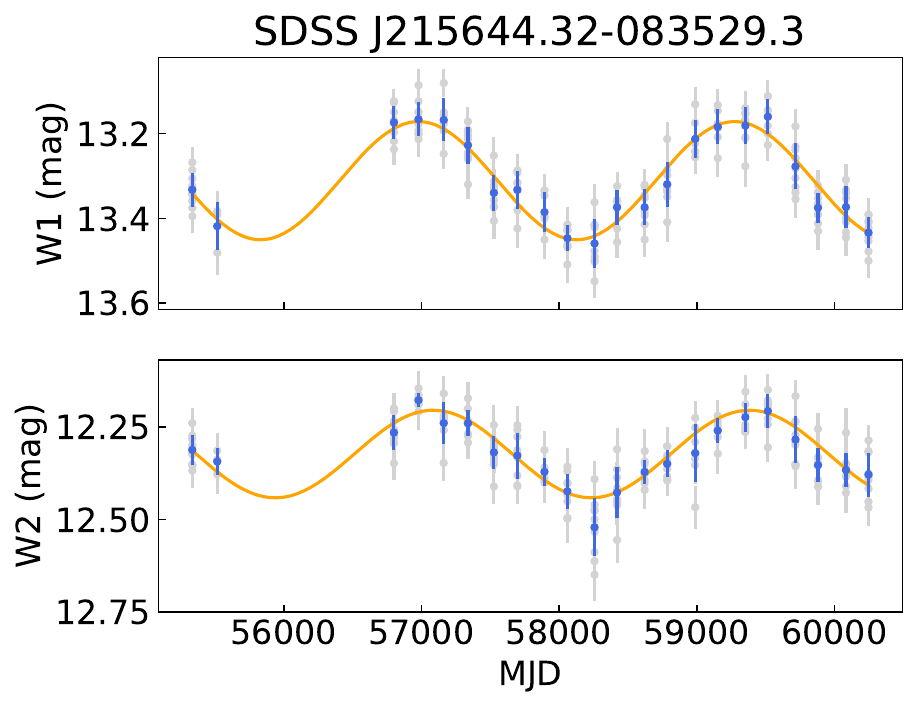}}
\centering{\includegraphics[width=0.49\textwidth]{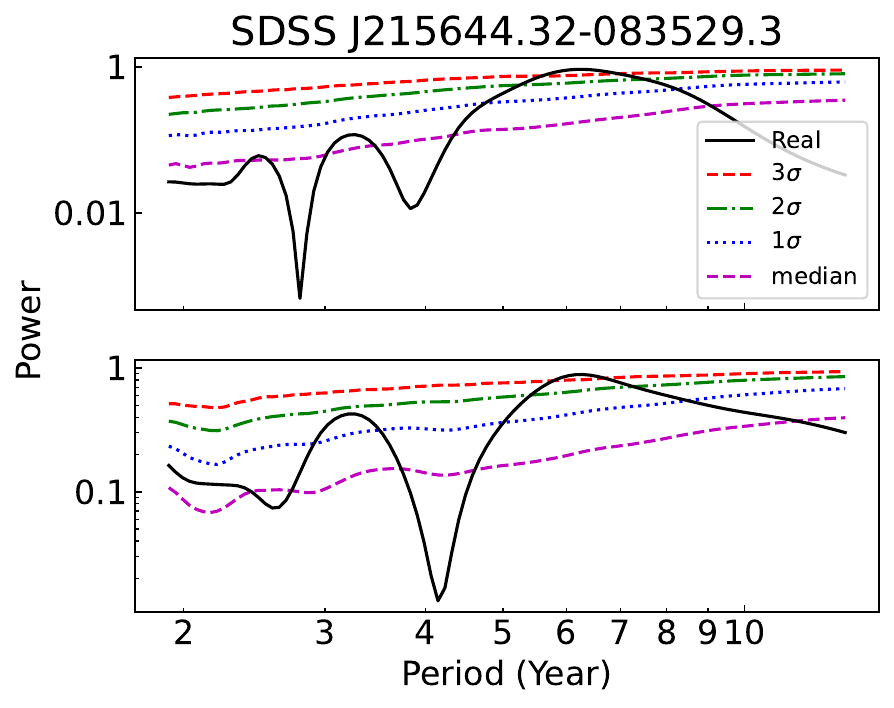}}
\end{minipage}
\begin{minipage}{1.0\textwidth}
\centering{\includegraphics[width=0.49\textwidth]{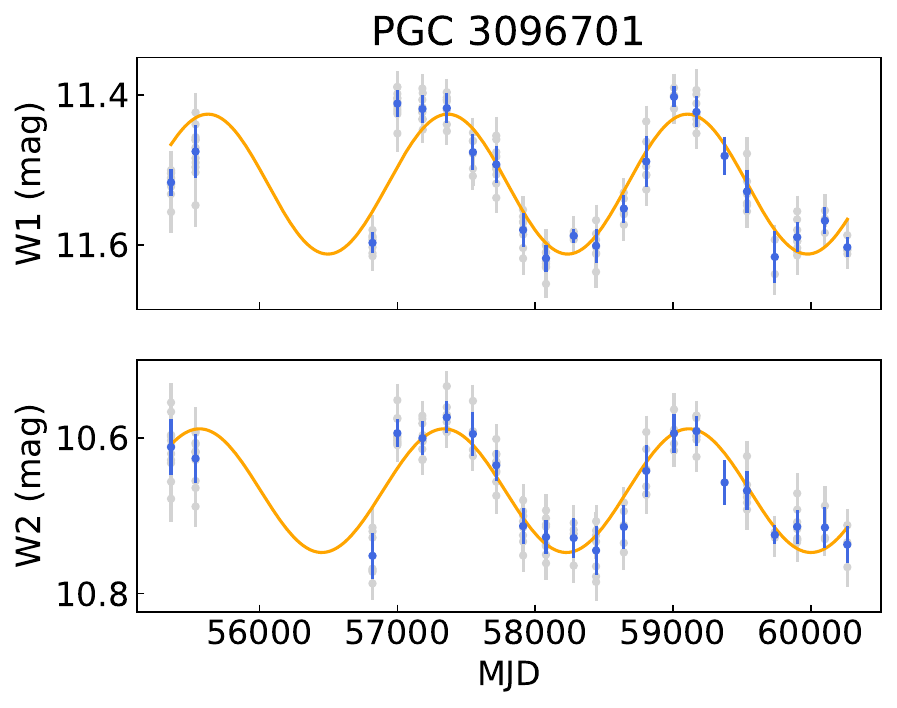}}
\centering{\includegraphics[width=0.49\textwidth]{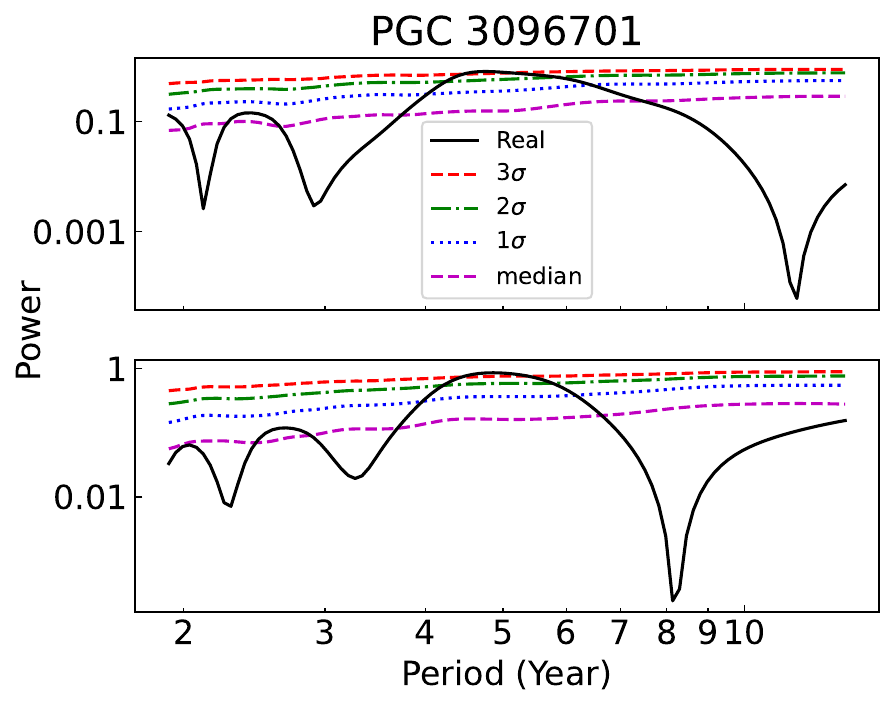}}
\end{minipage}
\addtocounter{figure}{-1}
\caption{(Continued).}
\end{figure*}

\end{appendices}

\end{document}